\documentstyle[12pt,epsfig,axodraw]{article}

\begin{document} 
\pagestyle{empty}

\begin{center} 
{\Large {\bf Short review about the MSSM with three right-handed neutrinos (MSSM3RHN).}}
\end{center}

\begin{center}
M. C. Rodriguez \\
{\it Grupo de F\'\i sica Te\'orica e Matem\'atica F\'\i sica \\
Departamento de F\'\i sica \\
Universidade Federal Rural do Rio de Janeiro - UFRRJ \\
BR 465 Km 7, 23890-000 \\
Serop\'edica, RJ, Brazil, \\
email: marcoscrodriguez@ufrrj.br \\} 
\end{center}

\begin{abstract}
We give a review about the Minimal Supersymmetric Standard Model with three right-handed neutrinos (MSSM3RHN). 
We, first introduce the minimal set of fields to built this model in their superfields formalism. After it, we build the 
lagrangian of the model in the superspace formalism and also introduce the soft terms to break SUSY. We show how to get 
masses to the neutrinos and sneutrinos in this model. We 
also present in short way some others interestings 
Supersymmetric model as 
SUSYB-L, SUSYLR and two SUSYGUTS models, all this model have three right handed neutrinos. 
\end{abstract}

PACS number(s): 12.60. Jv.

Keywords: Supersymmetric models.


\section{Introduction}

Although the Standard Model (SM) describes the observed properties
of charged leptons and quarks it is not the ultimate theory. 
However, the necessity to go beyond it, from the 
experimental point of view, comes at the moment only from neutrino 
data~\cite{bk}.

Today we have accumulated a variety of experimntal neutrino data, involving solar\footnote{The fluxes of neutrinos 
produced at the Sun, $\Phi(SSM)=6 \times 10^{10} \,\ \nu_{e} \,\ cm^{-2}s^{-1}$, was made at \cite{bahcall}.} (Homestake\footnote{This experiment was realized by Davis and his collaborators, in 1967, and it measured the electron neutrino flux produced from the Sun using a chlorine detector ($\nu_{e} +{}^{37}$Cl $\rightarrow$ $e^{-} +{}^{37}$Ar) in the Homestake mine (South Dakota)} \cite{RDavis}, Kamioka Nucleon Decay Experiment (Kamiokande) \cite{kam89, kam96}, Super-Kamiokande \cite{superK01}, Gallex \cite{Gallex},Gallium Neutrino Observatory (GNO) 
\cite{GNO}, Soviet American Gallium Experiment (SAGE) collaborations \cite{sage}), atmospheric (Irvine-Michigan-Brookhaven (IMB) \cite{imb86}, Kamioka Nucleon Decay Experiment (Kamiokande) \cite{kam_atm}), reactor (KamLAND \cite{kamland}, Borexino \cite{borexino}) and accelerator (K2K \cite{K2K},Minos \cite{MINOS}) experiments.
 
We know that the experiments Homestake mine 
(South Dakota, in USA), Kamiokande (Kamioka Nucleon Decay Experiment, in Japan), Super-Kamiokande experiment  provided the 
following main result:
\begin{itemize}
\item $\nu_{e}$ disappearance;
\end{itemize}
and it is know as ``solar neutrino puzzle" because 
those neutrinos were produced within the Sun. The 
neutrinos are emitted, basically, in the following 
``net" chain
\begin{equation}
4p+2e^{-}\rightarrow He^{4}+2 \nu_{e}+ \gamma .
\end{equation} 
and the fluxe of neutrinos produced at the Sun is 
\cite{bahcall}
\begin{equation}
\Phi(\nu_{e})=6 \times 10^{10} \,\ cm^{-2}s^{-1}.
\end{equation}

The solar neutrinos are detected by the following 
reaction
\begin{equation}
\nu_{x}+e \rightarrow \nu_{x}+e,
\end{equation}
where all the neutrinos contribute to this reaction. 
The result presented by Kamiokande is \cite{kam96}
\begin{eqnarray}
\Phi(\nu_{e})&=& \left( 2.80 \pm 0.019 (sta) \pm 0.33 (sys) \right) \times 10^{10}  \,\ cm^{-2}s^{-1}, 
\end{eqnarray}
while the results of Super-Kamiokande is
\begin{eqnarray}
\Phi(\nu_{e})&=& \left( 2.35 \pm 0.02 (sta) \pm 0.08 (sys) \right) \times 10^{10}  \,\ cm^{-2}s^{-1}.
\end{eqnarray}
Conclusion, the $\nu_{e}$ flux is depressed in 
relation with the fluxes of neutrinos 
produced at the Sun, it is about three times lower.

Another interesting data, is that the experiment Subdury Neutrino Observatory (SNO) \cite{SNO} has measured  the total amount of neutrinos $(\nu_{e},\nu_{\mu},\nu_{\tau})$ from the Sun agrees with the calculations of electron neutrinos production in the Sun, it means that, 
the flux of $(\nu_{\mu},\nu_{\tau})$ from the 
Sun is non-Zero and their results is
\begin{equation}
\frac{\Phi \left( \nu_{x} \right)}{\Phi(SSM)}= 
1 \pm 0.1,
\end{equation}
and we can interpretate it as the amount of 
neutrinos detected at SNO from the Sun 
is as expected, but the composition is not. Today it is the strongest experimental evidence for
\begin{itemize} 
\item neutrinos oscillation, 
\item existence of neutrinos masses.
\end{itemize}

While the IMB (Irvine-Michigan-Brookhaven) collaboration and again the Kamiokande experiment reported a significant deficit in the muon neutrino flux caming from the reaction 
$\pi^{-} \rightarrow \overline{\nu}_{\mu}\,\mu^{-} \rightarrow \overline{\nu}_{\mu}  \nu_{\mu} \nu_{e}\,e^{-}$
\footnote{We know when particles coming from the 
space\footnote{They are known as cosmic rays.} collide with nuclei in the upper atmosphere produce many pions.}, and it is know as atmospheric neutrino problem. The experimental results in this 
case can be summarized as
\begin{itemize}
\item The $\nu_{e}$ behave according to Montecarlo 
simulations;
\item There are a clear deficit in $\nu_{\mu}$.
\end{itemize}

The theory beyond oscillations can be summarize as: the neutrino state created in the decay 
\begin{equation}
W^{+}\rightarrow
l_{\alpha}^{+} + \nu_{\alpha},
\end{equation}
and it is described by the following lagrangian
\begin{equation}
{\cal L}_{W}= -
\frac{g}{\sqrt{2}} \sum_{\alpha=e, \mu, \tau} \left( \overline{l^{\prime}_{L\alpha}} \gamma^{m}  
\nu^{\prime}_{L\alpha} W_{m}^{-} + hc \right).
\label{eq2.1}
\end{equation}

In order to explain those series of data we taken into account that neutrinos flavor states 
\begin{eqnarray}
l^{\prime}_{L,R}&=& \left( 
e^{\prime},\mu^{\prime},\tau^{\prime}\right)^T_{L,R}, 
\nonumber \\
\nu^{\prime}_{L}&=&\left( 
\nu_{e},\nu_{\mu},\nu_{\tau}\right)^T_{L,R},
\end{eqnarray} 
and the massa eigenstates
\begin{eqnarray}
l_{L,R}&=& \left( 
e,\mu,\tau\right)^T_{L,R}, 
\nonumber \\
\nu_{L}&=&\left( 
\nu_{1},\nu_{2},\nu_{3}\right)^T_{L,R},
\end{eqnarray}
the mass matrices of charged lepton sector 
$M^{l}$ are diagonalized by a bi-unitary 
transformation such as 
\begin{equation}
\hat{M}^{l}=V^{l\dagger}_{L} M^{l} V^{l}_{R},
\end{equation} 
and 
\begin{equation}
\hat{M}^{l} = diag \left( m_{e}, m_{\mu}, m_{\tau} 
\right),
\end{equation} 
while the states $(\nu_{1},\nu_{2},\nu_{3})$ 
have definite masses given by $m_{1},m_{2},m_{3}$, 
respectively. We define the lepton mixing matrix as 
\begin{equation}
U_{PMNS}=V^{l\dagger}_{L}V^{\nu}_{L}.
\end{equation}

The unitary matrix, $U_{PMNS}$, is known as the Pontecorvo-Maki-Nakagawa-Sakata (PMNS). Then, Eq.(\ref{eq2.1}), become 
\begin{equation}
{\cal L}_{W}= -
\frac{g}{\sqrt{2}} \sum_{\alpha=e, \mu, \tau} 
\sum_{i=1,2,3} \left( \overline{l_{L\alpha}} 
\gamma^{m} U_{\alpha i} \nu_{Li} W_{m}^{-} + hc \right).
\label{eq2.2}
\end{equation}
Their experimental values are given by~\cite{GonzalezGarcia:2012sz}
\begin{equation}
\vert V_{PMNS}\vert \approx\left(\begin{array}{ccc}
0.795-0.846& 0.513-0.585 & 0.126-0.178\\
0.4205-0.543 & 0.416-0.730  & 0.579 - 0.808 \\
0.215 - 0.548 & 0.409 - 0.725 & 0.567 -0.800 \\
\end{array}\right),
\label{pmnsexp}
\end{equation}
he Majorana phases have not been included above 
because they do not lead to observable effects in 
oscillations experiments.

\begin{figure}
\centering
\includegraphics[width=0.8\textwidth]{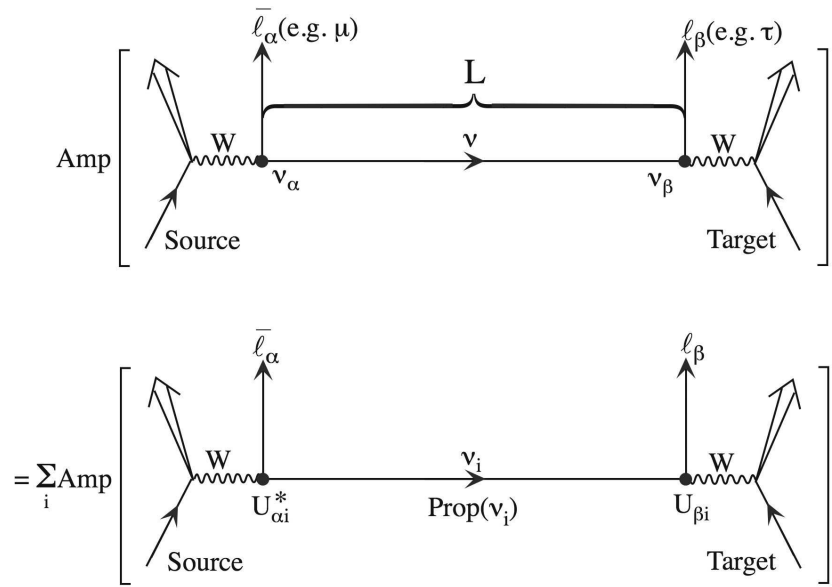}        
\caption{Neutrino flavor change (oscillation) 
in vacuum this figure was taken from 
\cite{Kayser:2005cd,Kayser:2008ev}.} 
\label{f1}
\end{figure}

The neutrino source, for example the Sun or some 
nuclear reaction, produce the neutrino 
$\nu_{\alpha}$ and its propageted until the detector, 
or target, but it arrive on it as in another flavor, 
as drawn at Fig.(\ref{f1}), $\nu_{\beta}$.

The amplitude for a neutrino of flavor $\nu_{\alpha}$
of energy $E$ changing to another neutrino flavor 
$\nu_{\beta}$ after propagating (in vacuum) a 
distance $L$ is given by
\begin{equation}
\langle \nu_{\beta}| \nu_{\alpha} \rangle
\simeq \sum_{i}\, U_{\alpha i}^{*}\, 
e^{- \imath \,m_{i}^{2} L / 2E}\, U_{\beta i}\;.
\label{eq1:osc_amp}
\end{equation}
We can calculate the corresponding probability of 
transition as:
\begin{eqnarray}
\mbox{Pr}(\nu_{\alpha}\rightarrow \nu_{\beta})
&=&
\delta_{\alpha\beta}
-4\,\sum_{i>j}\,
\mbox{Re}\left(U_{\alpha i}^*U_{\beta i}U_{\alpha j}U_{\beta j}^*\right)
\,\sin^2\left[\frac{\Delta m^2_{ij} \,L}{4E}\right] \nonumber\\
&+&
2\,\sum_{i>j}\,\mbox{Im}\left(U_{\alpha i}^*U_{\beta i}U_{\alpha j}U_{\beta j}^*\right)
\,\sin \left[\frac{\Delta m^2_{ij} \,L}{2E}\right]
\;, 
\label{eq1:osc_pr}
\end{eqnarray}
where we have defined 
\begin{equation}
\Delta m^{2}_{ij} \equiv 
m_{i}^{2}-m_{j}^{2}.
\end{equation}
The conclusion is clear, when all neutrino masses 
$m_{i}$'s are zero (or nonzero but degenerate), the 
neutrino oscillation is not possible and at least two 
of them must be massive.  

In similar way the transition $\overline{\nu_{\alpha}}\rightarrow 
\overline{\nu_{\beta}}$ is given by
\begin{eqnarray}
\mbox{Pr}(\overline{\nu_{\alpha}}\rightarrow 
\overline{\nu_{\beta}})
&=&
\delta_{\alpha\beta}
-4\,\sum_{i>j}\,
\mbox{Re}\left(U_{\alpha i}^*U_{\beta i}U_{\alpha j}U_{\beta j}^*\right)
\,\sin^2\left[\frac{\Delta m^2_{ij} \,L}{4E}\right] \nonumber\\
&-&
2\,\sum_{i>j}\,\mbox{Im}\left(U_{\alpha i}^*U_{\beta i}U_{\alpha j}U_{\beta j}^*\right)
\,\sin \left[\frac{\Delta m^2_{ij} \,L}{2E}\right]
\;, 
\label{eq2:osc_pr}
\end{eqnarray}
and as conclusion if the matrix $U$ is not real, 
the probabilities for 
$\nu_{\alpha}\rightarrow \nu_{\beta}$ and for the 
corresponding antineutrino oscillation, 
$\overline{\nu_{\alpha}}\rightarrow 
\overline{\nu_{\beta}}$, will 
in general differ.

As we discuss above, we must have two masses scales, 
and there are the following free parameters:
\begin{itemize}
\item three mixing angles ($\theta_{12},\theta_{23}$ and $\theta_{13}$);
\item the $CP$ violating Dirac\footnote{We meant Dirac 
particle \cite{Dirac}.} phase ($\delta$);
\item two Majorana\footnote{We meant Majorana 
particle \cite{Majorana}.} phases ($\alpha_{1}$  and $\alpha_{2}$);
\item two squared-mass differences (e.g. $\Delta m_{21}^{2}\equiv m^{2}_{2}-m^{2}_{1}$ and $\Delta m_{31}^{2}\equiv m^{2}_{3}-m^{2}_{1}$).
\end{itemize}

The unitary matrix, $U_{PMNS}$, is often parametrized as
\begin{eqnarray}
U_{PMNS} &=& 
\left(
\begin{array}{ccc}
c_{12} & s_{12} & 0 \\
-s_{12} & c_{12} & 0 \\
0 & 0 & 1
\end{array}
\right)
\left(
\begin{array}{ccc}
c_{13} & 0 & s_{13}e^{- \imath \delta} \\
0 & 1 & 0 \\
-s_{13}e^{\imath \delta} & 0 & c_{13}
\end{array}
\right)
\left(
\begin{array}{ccc}
1 & 0 & 0 \\
0 & c_{23} & s_{23} \\
0 & -s_{23} & c_{23}
\end{array}
\right) \nonumber \\
&\times&
\left(
\begin{array}{ccc}
e^{\imath (\alpha_{1}/2)} & 
0 & 0 \\
0 & 
e^{\imath (\alpha_{2}/2)} & 0 \\
0 & 0 & 1
\end{array}
\right)
, \nonumber \\
\label{eq1:UPMNS_param} 
\end{eqnarray}
where as usual, we have defined 
\begin{eqnarray}
s_{mn}&\equiv& \sin \theta_{mn}, \nonumber \\ 
c_{mn}&\equiv& \cos \theta_{mn}.
\end{eqnarray}

\begin{figure}
\centering
\includegraphics[width=0.8\textwidth]{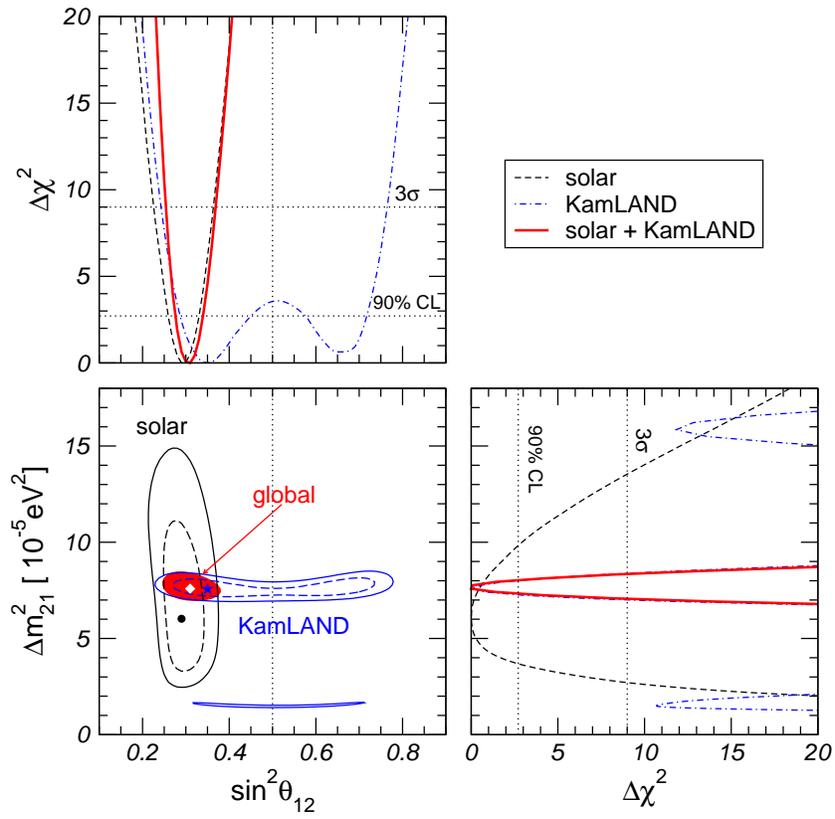}
\caption{Determination of the leading ``solar'' 
oscillation parameters, taken from \cite{nu_best-fit}.}
\label{fig:dominant-12}
\end{figure}

\begin{figure}
\centering
\includegraphics[width=0.8\textwidth]{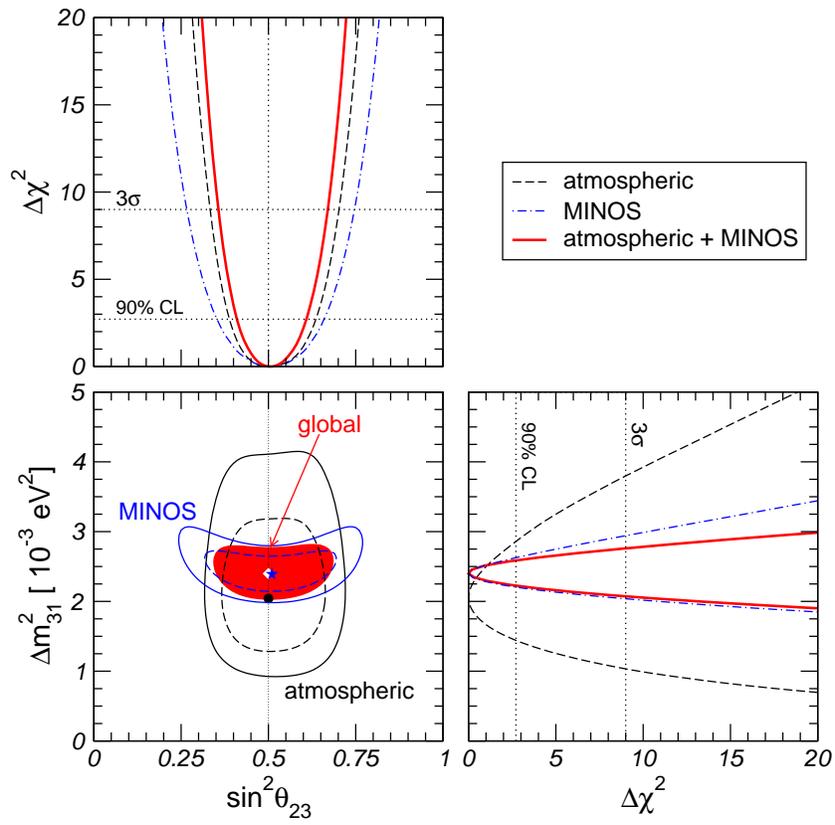}
\caption{Determination of the leading 
``atmospheric'' oscillation parameters, taken 
from \cite{nu_best-fit}.}
\label{fig:dominant-23}
\end{figure}

The experimental results that suggest
that the neutrinos have non-zero masses and oscillations. The best-fit values at $1 \sigma$ error level for these neutrino oscillation parameters in the three-flavor framework are summarised as 
follows \cite{nu_best-fit}
\begin{eqnarray}
\sin^{2}\theta_{12}&=&\sin^{2} \theta_{solar} = 0.304^{+0.022}_{-0.016} \,,
\Delta m^{2}_{21}= \Delta m_{solar}^{2} = 7.65^{+0.23}_{-0.20} \times 10^{-5}\,\ {\mbox eV}^{2}\,\ Fig.(\ref{fig:dominant-12}) \;, \nonumber \\
\sin^{2} \theta_{23} &=& \sin^{2} \theta_{atm} = 0.50^{+0.07}_{-0.06}\,,
|\Delta m^{2}_{23}|= \Delta |m_{atm}^{2}| = 2.40^{+0.12}_{-0.11} \times 10^{-3}\,\ {\mbox eV}^{2}\,\ 
Fig.(\ref{fig:dominant-23}) \;, \nonumber \\
\sin^{2} \theta_{13} &=&\sin^{2} \theta_{CHOOZ} =  0.01^{+0.016}_{-0.011}\;. \nonumber \\
\label{eq1:best-fit_mass}
\end{eqnarray}
From these data we can conclude
\begin{itemize}
\item The mixing $\theta_{23}$ is consistent with 
maximal mixing;
\item The mixing $\theta_{12}$ is large but not maxima;
\item The CHOOZ \cite{chooz} (in France) results 
indicate a tiny value for the mixing angle 
$\theta_{13}$.
\end{itemize}

We can consider two scenarios\footnote{Here we have 
omitted the Majorana phases because they do not lead 
to observable effects in oscillations.}:
\begin{itemize}
\item ``bi-large'' mixing, where the mixing 
parameters are $\theta_{23} = \pi/4$ and 
$\theta_{13}=0$
\begin{equation}
U_{BL} = 
\left(
\begin{array}{ccc}
\cos\theta_{12} & \sin\theta_{12} & 0 \\
-\frac{\sin\theta_{12}}{\sqrt{2}} & 
\frac{\cos\theta_{12}}{\sqrt{2}} & 
\frac{1}{\sqrt{2}} \\
-\frac{\sin\theta_{12}}{\sqrt{2}} & 
\frac{\cos\theta_{12}}{\sqrt{2}} & 
-\frac{1}{\sqrt{2}}
\end{array}
\right) 
\label{eq1:osc_BiLarge}
\end{equation}
\item ``tribimaximal'' mixing, where the mixing 
parameters are $\theta_{12}$ is very well approximated by the relation: $\sin^2 \theta_{12} = 1/3$ and te 
Eq.(\ref{eq1:osc_BiLarge}) can be written as
\begin{equation}
U_{TB} = 
\left(
\begin{array}{ccc}
\sqrt{\frac{2}{3}} & \frac{1}{\sqrt{3}} & 0 \\
-\frac{1}{\sqrt{6}} & 
\frac{1}{\sqrt{3}} & \frac{1}{\sqrt{2}} \\
-\frac{1}{\sqrt{6}} & 
\frac{1}{\sqrt{3}} & -\frac{1}{\sqrt{2}}
\end{array}
\right) 
\label{eq1:osc_TriBi}
\end{equation} 
\end{itemize}

\begin{figure}[!tbp]
\begin{center}
\hspace{-0.1cm} 
\epsfxsize9cm\epsffile{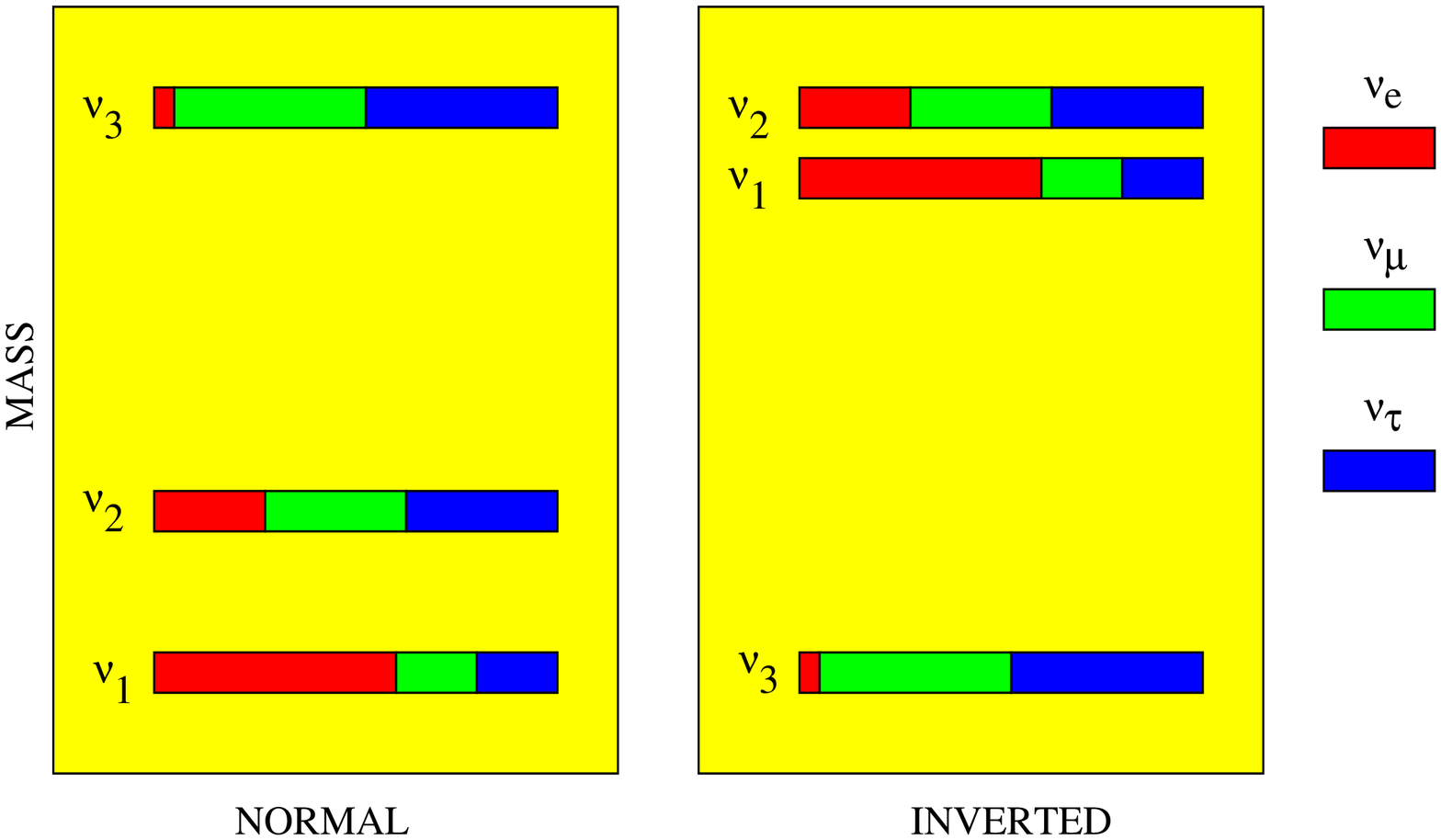} 
\caption{Neutrino
mass and flavor spectra for the normal (left) and inverted (right)
mass hierarchies, figure taken from 
\cite{Mohapatra:2006gs}. }
\label{sp}
\end{center}
\end{figure}

In the case of atmospheric neutrinos, the sign of 
the mass split determines the type
of mass hierarchy:  
\begin{itemize}
\item normal, $\Delta m_{32}^2 > 0$ wchich gives 
\begin{equation}
m_{1}<m_{2}<m_{3},
\end{equation}
with
\begin{equation}
m_{2} = \sqrt{m_{1}^{2} +\Delta m_{solar}^{2}}, 
\,\ 
m_{3} = \sqrt{m_{1}^2 +\Delta m_{atm}^{2}}\;, 
\label{eq1:NH_m2_m3}
\end{equation}
\item inverted, 
$\Delta m_{32}^2 <0 $
\end{itemize} 
wchich gives 
\begin{equation}
m_{3}<m_{1}<m_{2},
\end{equation}
\begin{equation}
m_{1} = \sqrt{m_{3}^{2} + 
\Delta m_{atm}^{2}- \Delta m_{solar}^{2}},
\,\ 
m_{2} = \sqrt{m_{1}^{2} +\Delta m_{solar}^{2}}, 
\label{eq1:IH_m1_m2}
\end{equation}
and both situation are draw at Fig.(\ref{sp}). To explain the data presented above, we need three massive neutrinos. In order to see how to fit these data with three massive neutrinos see 
\cite{Mohapatra:2006gs,nu_best-fit}.

The physical effects involved in the interpretation 
of neutrinos oscillation data are \cite{Mohapatra:2006gs}
\begin{itemize}
\item vacuum oscillations;
\item the Mikheyev-Smirnov-Wolfenstein (MSW) (conversion
of  solar neutrinos in the matter of the Sun);
\item oscillations in matter (oscillations of solar and atmospheric
neutrinos in the matter of Earth).
\end{itemize}

The Los Alomos Liquid Scintillation Detector 
experiment (LSND) measured accelerator neutrinos 
produced in the usual muon\'s decays
\begin{equation}
\mu^{+}\rightarrow e^{+}\nu_{e}\bar{\nu}_{\mu}.
\end{equation}
They have reported a signal that, if interpreted as 
neutrino oscillations, implies \cite{Athanassopoulos:1996jb}
\begin{equation}
\Delta m^{2}\sim {\cal O} (1\,{\mbox{\rm eV}}^{2}),
\end{equation} 
and the mixing parameter is constrained by
\begin{equation}
0.003 \leq sin^{2} \left( 2 \theta_{LSND} \right) 
\leq 0.03.
\end{equation}  
It is important to remeber that this results has not 
been confirmed by any other experiment. Due this 
fact, usually their 
results are ignored because an oscillation 
explanation has been ruled out by MiniBooNE 
\cite{MiniBooNE}. If we 
take into account their result, we need to introduce 
at least one right-handed neutrinos\footnote{They are known as sterile neutrinos, for more detail about those kind of names see \cite{Volkas:2001zb}.}.

About the neutrino physics, one of the major problem is to clarify the Dirac 
or Majorana character of neutrino masses. If neutrinos are Dirac particles 
then lepton number remains as a conserved quantity. By another hand if 
neutrinos are Majorana particles \cite{Pal:2010ih} lepton number would be violated in two units. In this 
case double beta decays and some rare meson decay of the 
form $M^{+} \to M^{\prime -}l^{+}l^{+}$ can occur. This kind of decay is 
sensitive to neutrino masses and lepton mixing as discussed at \cite{cvetic}. 

\begin{figure}[!tbp]
\begin{center}
\hspace{-0.1cm} 
\epsfxsize9cm\epsffile{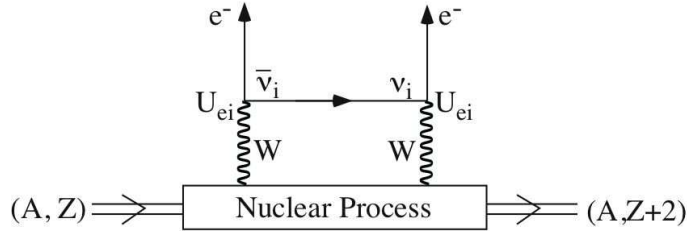} 
\caption{The dominant mechanism for $0\nu\beta\beta$, 
this figure taken from \cite{Kayser:2008rd}.}
\label{f4.1}
\end{center}
\end{figure}

One possible way to probe the neutrino mass scale 
is to measures the neutrinoless double $\beta$-decay. 
This process occur via the following nuclear reaction
\begin{equation}
^{A}_{Z}\left[\mbox{Nucl}\right] 
\rightarrow \;^{\;\;\;\;A}_{Z+2}
\left[\mbox{Nucl}^{\prime}\right] + 2 e^{-},
\end{equation}
this equation is translated as a nucleus containing 
$Z$ protons decays to a nucleus containing $Z+2$ 
protons by emitting two electrons, 
an example is $^{76}Ge \rightarrow ^{76}Se$, as 
drawn at Fig.(\ref{f4.1}).

However, several groups such as the 
Heidelberg-Moscow \cite{heidelberg-moscow}, IGEX \cite{IGEX} and  CUORICINO 
experiment \cite{CUORICINO}. However there are the 
following upcoming experiments like CUORE \cite{CUORE}, GERDA \cite{GERDA} and \textsl{Majorana} \cite{MAJORANA} to get better 
limits on neutrinos masses. 

If we take into account cosmological constriant into account we get the following upper limit 
$\sum m_{\nu}<1.01$ eV if we have three neutrinos and it change to $\sum m_{\nu}<2.12$ eV in the case of five neutrinos as presented at 
\cite{Hannestad:2003xv,Hannestad:2003jt}.

The aim of the present paper is to propose a modification Minimal Supersymmetric Standard Model to give masses in a simple way to neutrinos and get 
nice Dark Matter candidate (right-handed sneutrinos) and also get flat directions scalar potential that can generate the cosmological inflation. 
Also, the matter anti-matter asymmetry could be obtained from the letogenesis mechanism. 

I started this studies to make the articles presented at 
\cite{Rodriguez:2016esw,Montero:2016qpx,Rodriguez:2020fvo,axionneutrinodm,global,local}, my first goal was to include it on my previous review about some generalization about the MSSM 
with singlets \cite{Rodriguez:2019mwf}, but this model is so interesting due it I have done this review. 

This paper is organized as follows. In Section \ref{sec:mssm3rhn} we present the minimal supersymmetric model with three right-handed neutrinos. We calculate  the masses to neutrinos 
and sneutrinos in general case with three 
right-handed neutrinos and also with with one 
right-handed neutrinos, as an example as of that 
this model can explain in satisfatory way the 
experimental results concern the neutrino\'s 
experimental data. Next, 
we show some flat directions in this model. We also present, very quickly, some others interestings supersymmetric models such as SUSYB-L, SUSYLR and two SUSYGUTS models. 

\section{The Minimal Supersymmetric Model with three right-handed neutrinos (MSSM3RHN)}
\label{sec:mssm3rhn}

On the Minimal Supersymmetric Model with three right-handed neutrinos (MSSM3RHN) \cite{Baer:2006rs,Pagetese,f3,5,6}, the particle content is very similar 
to the Minimal Supersymmetric Standard Model (MSSM) \cite{Baer:2006rs,dress}, it contains three families of left-handed quarks $Q_{iL}=(u_{i},d_{i})^{T}_{L}\sim \left( {\bf 3}, {\bf 2}, (1/3) \right)$\footnote{The number in parenthesis refers to the $(SU(3)_{C},SU(2)_{L},U(1)_{Y})$ quantum numbers.}, $i=1,2,3$ is familly 
indices, three families of leptons 
$L_{iL}=(\nu_{i},l_{i})^{T}_{L}\sim \left( {\bf 1}, {\bf 2},-1 \right)$ plus the Higgs fields $H_{1}=(h^{+}_{1},h^{0}_{1})^{T}\sim \left( {\bf 1}, {\bf 2},1 \right)$. We have also to introduce three families of right-handed quarks, given by  
$(u_{iR}\sim \left( {\bf 3}^{*}, {\bf 1}, (-4/3) \right)$, $d_{iR}\sim \left( {\bf 3}^{*}, {\bf 1}, (2/3) \right)$)\footnote{We are using similar notation as presented at \cite{Rodriguez:2019mwf}.}, three families 
of right-handed charged leptons 
($l_{iR}\sim \left( {\bf 1}, {\bf 1},2 \right)$) 
and another Higgs fields 
$H_{2}=( h^{0}_{2}, h^{-}_{2})^{T}\sim \left( {\bf 1}, {\bf 2},-1 \right)$ as shown at Tab.(\ref{tab:mssm3rhn}). The vaccum expectation values 
(vev) of our Higgses fields are given by:
\begin{eqnarray}
\langle H_{1} \rangle = \frac{1}{\sqrt{2}} \left(
\begin{array}{c}
0 \\
v_{1} 
\end{array}
\right), \,\
\langle H_{2} \rangle = \frac{1}{\sqrt{2}} \left(
\begin{array}{c}
v_{2} \\
0 
\end{array}
\right).
\label{vevhiggs}
\end{eqnarray} 

\begin{table}[t]
\center
\renewcommand{\arraystretch}{1.5}
\begin{tabular}
[c]{|l|cc|cc|}\hline
Superfield & Usual Particle & Spin & Superpartner & Spin\\
\hline
\hline
\quad 
$\hat{L}_{iL}\sim({\bf 1},{\bf 2},-1)$ & $L_{iL}=(\nu_{iL},\,l_{iL})^{T}$ & 
$\frac{1}{2}$ & $\tilde{L}_{iL}=(\tilde{\nu}_{iL},\,\tilde{l}_{iL})^{T}$ & 0\\
\quad
$\hat{Q}_{iL}\sim({\bf 3},{\bf 2},1/3)$ & $Q_{iL}=(u_{iL},\,d_{iL})^{T}$ & $\frac
{1}{2}$ & $\tilde{Q}_{iL}=(\tilde{u}_{iL},\,\tilde{d}_{iL})^{T}$ & 0\\
\hline
\quad 
$\hat{l}_{iR}\sim({\bf 1},{\bf 1},2)$ & $l_{iR}= \overline{l^{c}}_{iL}$ & 
$\frac{1}{2}$ & $\tilde{l}_{iR}$ & 0\\
\quad 
$\hat{N}_{iR}\sim({\bf 1},{\bf 1},0)$ & $\nu_{iR}= \overline{\nu^{c}}_{iL}$ & 
$\frac{1}{2}$ & $\tilde{\nu}_{iR}$ & 0\\
\quad 
$\hat{d}_{iR}\sim({\bf 3^{\ast}},{\bf 1},(2/3))$ & $d_{iR}= \overline{d^{c}}_{iL}$ &
$\frac{1}{2}$ & $\tilde{d}^{c}_{iL}$ & 0\\
\quad 
$\hat{u}_{iR}\sim({\bf 3^{\ast}},{\bf 1},(-4/3)$ & $u_{iR}= \overline{u^{c}}_{iL}$ &
$\frac{1}{2}$ & $\tilde{u}^{c}_{iL}$ & 0\\
\hline
\quad 
$\hat{H}_{1}\sim({\bf 1},{\bf 2},1)$ & 
$H_{1}=(h_{1}^{+},\, h_{1}^{0})^{T}$ & 0 &
$\tilde{H}_{1}=(\tilde{h}_{1}^{+},\, \tilde{h}_{1}^{0})^{T}$ & $\frac{1}{2}$\\
\quad 
$\hat{H}_{2}\sim({\bf 1},{\bf 2},-1)$ & 
$H_{2}=(h_{2}^{0},\, h_{2}^{-})^{T}$ & 0 &
$\tilde{H}_{2}=(\tilde{h}_{2}^{0},\, \tilde{h}_{2}^{-})^{T}$ & $\frac{1}{2}$\\
\hline
\quad 
$\hat{V}^{a}_{c} (SU(3))$ & $G^{a}_{m}$ & 1 & $\tilde{ g}^{a}$ &
$\frac{1}{2}$\\
\quad 
$\hat{V}^{i}$ (SU(2)) & $V^{i}_{m}$ & 1 & $\lambda^{i}_{A}$ & $\frac
{1}{2}$\\
\quad 
$\hat{V}^{\prime}$ (U(1)) & $V_{m}$ & 1 & $\lambda_{B}\,\,$ & $\frac
{1}{2}$\\
\hline
\end{tabular}
\renewcommand{\arraystretch}{1}\caption{Particle content of MSSM3RHN. The  
families index for leptons and quarks are $i,j=1,2,3$. The parentheses are the transformation 
properties under the respective representation of $(SU(3)_{C},SU(2)_{L},U(1)_{Y})$.}
\label{tab:mssm3rhn}
\end{table}

The superpotential of the MSSM is define in the following way
\begin{equation}
W_{MSSM}= \left( \frac{W_{H}+W_{2RV}}{2} \right) + \left( \frac{W_{Y}+W_{3RV}}{3} \right) +hc,
\label{mssmsuppotrc}
\end{equation}
where
\begin{eqnarray}
W_{H}&=& \mu_{H} \left( \hat{H}_{1}\hat{H}_{2} \right), \nonumber \\ 
W_{Y}&=&\sum_{i,j=1}^{3}\left[\,
f^{l}_{ij}\left( \hat{H}_{2}\hat{L}_{iL} \right) \hat{l}_{jR}+
f^{d}_{ij}\left( \hat{H}_{2}\hat{Q}_{iL} \right) \hat{d}_{jR}+
f^{u}_{ij}\left( \hat{H}_{1}\hat{Q}_{iL} \right) \hat{u}_{jR}\,\right], 
\label{Y-H}               
\end{eqnarray}
we have suppressed the $SU(2)$ indices and the supersymmetric  parameter $\mu_{H}$ is a complex number 
and has mass dimension. In general all the parameters $f$ are, in principle, complex numbers and they are symmetric in $ij$ exchange and they are 
dimensionless parameters \cite{Baer:2006rs,dress}. Moreover, $f^{d}$ and $f^{u}$ can give account for the mixing between the quark current eigenstates 
as described by the CKM matrix. In this model, we can also explain the mass hierarchy in the charged fermion masses as showed recently in 
\cite{cmmc,cmmc1}.  

When we consider $R$ Parity scenarios, this model has viable candidates 
to be the Dark Matter. One of most studied candidate is the lightest 
neutralino denoted as $\tilde{\chi}^{0}_{1}$. However, some years ago, 
there were some intersting studies where the lighest sneutrino 
$\tilde{\nu}_{eL}$ was considered as the Lighest Supersymmetric Particle 
(LSP)\footnote{Unfortunatelly this particle have been ruled out by the combination of collider experiment as LEP and direct 
searches for cosmological relics as discussed at \cite{hillwalkers}.}. There are some authors study the gravitino as Dark Matter candidate, more 
details about SUSY Dark Matter Candidates is presented in nice way in \cite{Ellis:2010kf}.

The MSSM has terms to break the baryon number and the lepton number conservation laws 
and in the SM all the interactions conserve both laws and no physical process with this property has been discovered so far. This phenomenological 
fact suggest imposing a discrete symmetry in the model. This symmetry is the $R$-parity \cite{Baer:2006rs,dress}. The terms that break the $R$-parity 
are given by
\begin{eqnarray}
W_{2RV}&=&\sum_{i=1}^{3}\mu_{i} \left( \hat{H}_{1}\hat{L}_{iL} \right),\nonumber \\
W_{3RV}&=&\sum_{i,j,k=1}^{3} \left[  
\lambda_{ijk} \left( \hat{L}_{iL}\hat{L}_{jL}\right) \hat{l}_{kR}+
\lambda^{\prime}_{ijk} \left( \hat{L}_{iL}\hat{Q}_{jL}\right) \hat{d}_{kR}+ 
\lambda^{\prime\prime}_{ijk}\hat{u}_{iR}\hat{d}_{jR}\hat{d}_{kR} \right] . \nonumber \\
\label{mssmrpv}
\end{eqnarray}

\begin{figure}[ht]
\centerline{\epsfxsize=3.1in\epsfbox{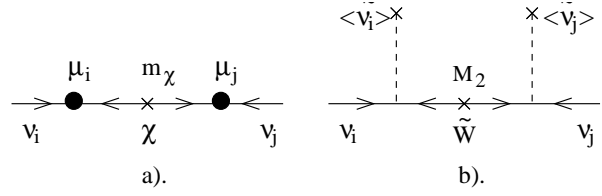}}   
\caption{Diagrams for the neutrino mass generation at tree level in the 
model with the R-parity violation figure taken from \cite{Smirnov:2004hs}.   
\label{rpvt}}
\end{figure}

\begin{figure}[ht]
\begin{center}
\vglue -0.009cm 
\mbox{\epsfig{file=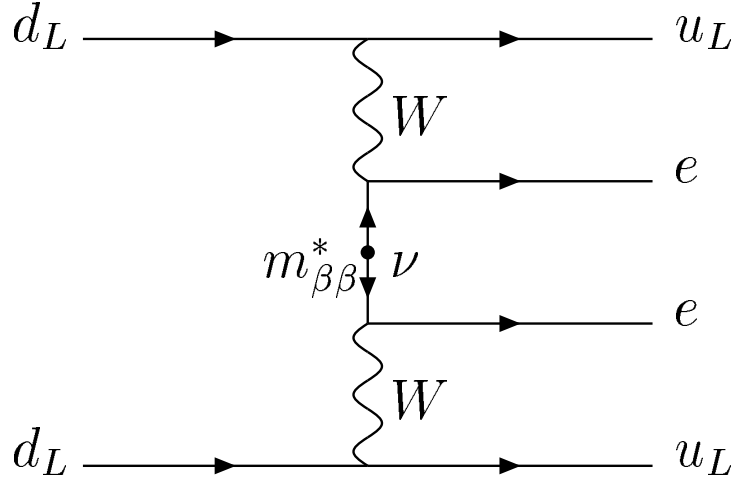,width=0.7\textwidth,angle=0}}
\end{center}
\caption{Neutrinoless double beta  decay in  SM with massive
neutrinos.} 
\label{ddbsnmp}
\end{figure}

\begin{figure}[ht]
\begin{center}
\vglue -0.009cm 
\mbox{\epsfig{file=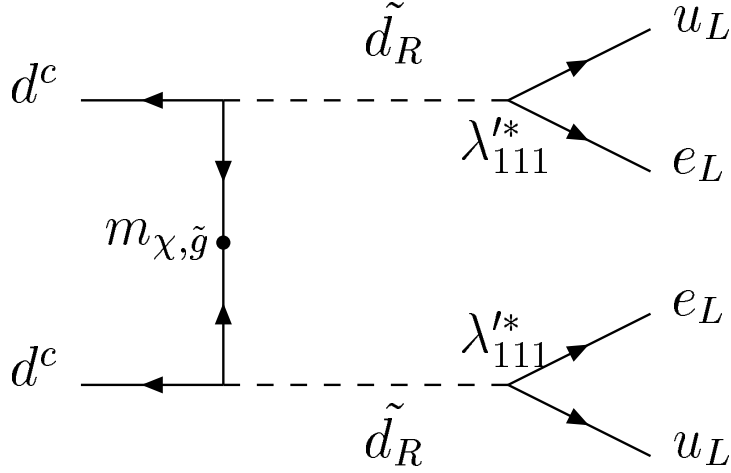,width=0.7\textwidth,angle=0}}
\end{center}
\caption{Neutrinoless double beta ($0\nu\beta\beta$) decay in the
MSSM with massive neutrinos.} \label{ddbsnmpsm}
\end{figure}

The mass matrix of neutrinos arise when we allow a mixing between the usual leptons with the higgsinos and its mixings is generated by
\begin{equation} 
\left( \hat{H}_{2}\hat{L}_{iL} \right) \supset \left( \tilde{H}_{2}L_{iL} \right) =l_{iL}\tilde{H}^{+}_{2}- \nu_{iL}\tilde{H}^{0}_{2},
\end{equation}
the second term at right size mixing the neutrinos with the neutral higgsinos, therefore in this case the usual neutrinos can mix with 
the neutralinos, and one of neutrinos get mass at Tree-level as shown at Fig.(\ref{rpvt}). The others two neutrinos get their masses at 1-loop 
level, neutrinos are Majorana particles, and their masses at can be written as \cite{dress}:
\begin{equation}
\delta m_{\nu_k}\propto 
\left( \lambda_{kii}\, {\mbox or}\, \lambda^{\prime}_{kii} \right)^2
\frac{M_Sm^2_i}{\tilde{m}_i},
\label{numass1}
\end{equation}
where $m_i$ is the mass of exchanged fermion and the factor 
$m_{i}M_{S}$ comes from the left-right mixing of the sfermion, we 
recomend to see 
\cite{banks,hall,Romao:1991ex,rv1,Davidson:2000ne,Montero:2001ch,
Smirnov:2004hs}. We can have neutrinoless double beta decay ($0\nu\beta\beta$), see \cite{Rodriguez:2016esw}, as at Figs.(\ref{ddbsnmp},\ref{ddbsnmpsm}).

As we broke $L$-number conservation we can have Leptogenesis  
\cite{yanagida,Law:2009vh,Law:2010zz} because on this case the 
following decay:
\begin{equation}
\Gamma \left( 
\tilde{\nu}_{i} \to l^{+}_{j}l^{-}_{k}
\right) = \frac{1}{16 \pi}(\lambda_{ijk})^2m_{\tilde{\nu}_i},
\label{fcnc}
\end{equation}
where $m_{\tilde{\nu}_i}$ is the mass of sneutrinos and this 
decay violate lepton number conservation. Process violating Lepton 
number can be confirmed at LHC, their detectors can get the following 
processes
\begin{eqnarray}
pp & \rightarrow & \tilde{e}_{i} \rightarrow e^{-} \tilde{\chi}_{1}^{0} \rightarrow e^+e^-jj, \nonumber \\
pp & \rightarrow & \tilde{\nu_{eL}} \rightarrow e^{-} \tilde{\chi}_{1}^{+} \rightarrow e^+e^-jj.
\label{cmsexplanation}
\end{eqnarray}
and its total cross section is give by
\begin{eqnarray}
\sigma (pp \rightarrow \tilde{e}) \propto |\lambda^{\prime}_{111}|^{2} / m_{\tilde{e}}^{3}, 
\end{eqnarray}
and those processes mentioned above can be confirmed by ATLAS, CMS and 
LHCb and it provides an attractive scenario to explain the baryon 
asymmetry \cite{Rodriguez:2016esw}.

When we consider $\tilde{\chi}^{0}_{1}$ as the LSP, in scenarios where $R$-Parity is conserved, due the 
terms at Eq.(\ref{mssmrpv}) induces the following decays channel:
\begin{itemize}
\item $\lambda_{ijk}\,\ \Longrightarrow \,\ 
\tilde{\chi}^{0}_{1} \to l_{i}\bar{l}_{j}\bar{\nu}_{k}, \,\ {\mbox or} \,\
\tilde{\chi}^{0}_{1} \to \bar{l}_{i}l_{j}\nu_{k},$
\item $\lambda^{\prime}_{ijk}\,\ \Longrightarrow \,\ 
\tilde{\chi}^{0}_{1} \to \bar{l}_{i}\bar{u}_{j}d_{k}, \,\ {\mbox or} \,\
\tilde{\chi}^{0}_{1} \to l_{i}u_{j}\bar{d}_{k},$
\item $\lambda^{\prime \prime}_{ijk}\,\ \Longrightarrow \,\
\tilde{\chi}^{0}_{1} \to \bar{u}_{i}\bar{d}_{j}\bar{d}_{k}, \,\ 
{\mbox or} \,\
\tilde{\chi}^{0}_{1} \to u_{i}d_{j}d_{k},$
\end{itemize}
and same on this case we can get the famous missing energy and momentum 
in MSSM, for more details see~\cite{Baer:2006rs,dress}.

\begin{figure}[ht]
\centerline{\epsfxsize=3.1in\epsfbox{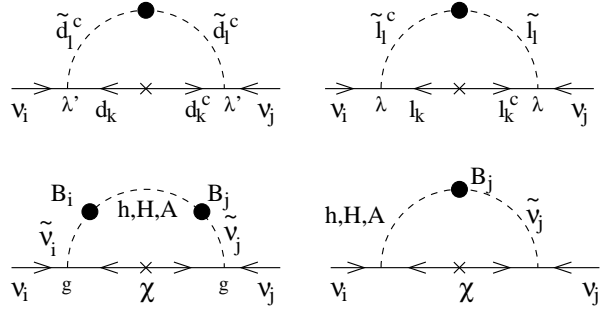}}   
\caption{One loop diagrams for the neutrino mass generation in the model 
with R-parity violation, figure taken from \cite{Smirnov:2004hs}.
\label{rpvl}}
\end{figure}

The main motivation to consider the MSSM with three right-handed 
neutrinos (MSSM3RHN) is to get an simple way to generate masses to all 
neutrinos in concordance with the actual experimental data. We add to MSSM 
particle spectra, three additional scalar superfields $\hat{N}_{1,2,3}$, see Tab.(\ref{tab:mssm3rhn})\footnote{In chiral superfields all the fermion fields are left-handed \cite{Rodriguez:2016esw}} 
\begin{eqnarray}
\hat{N}_{iR}(y, \theta )&=& \tilde{\nu}^{c}_{iL}(y)+ \sqrt{2} \left( \theta \nu^{c}_{iL}(y) \right) + 
\left( \theta \theta \right) F_{\nu^{c}_{iL}}(y), 
\end{eqnarray}
the fields $\nu^{c}_{iL}$ are the right-handed neutrinos, known as sterile neutrinos \cite{Volkas:2001zb}. As a consequence 
of supersymmetric algebra beyond the three right-handed neutrinos $\nu^{c}_{iL}\equiv \nu_{iR}$, or sterile neutrinos, we should 
also introduce three right-handed sneutrnios $\tilde{\nu}^{c}_{iL}$, and we will call them as sterile sneutrinos and they are 
good dark matter candidate \cite{Pagetese}. The new coordinate $y$ is defined as \cite{wb}
\begin{eqnarray}
y^{m}&=&x^{m}+ \imath \left( \theta \sigma^{m} \bar{\theta} \right).
\label{fermioniccoordinate}
\end{eqnarray}
The particle content of each this model is presented in the Tab.(\ref{tab:mssm3rhn}).  

The Lagrangian of this model is written as
\begin{equation}
{\cal L}_{MSSM3RHN}={\cal L}_{SUSY}+{\cal L}_{soft}\,\ ,
\label{totallagmssm3rhn}
\end{equation}
where ${\cal L}_{SUSY}$ is the supersymmetric piece and can be divided as follows
\begin{equation}
{\cal L}_{SUSY}={\cal L}_{lepton}+{\cal L}_{Quarks}+{\cal L}_{Gauge}+{\cal L}_{Higgs},
\end{equation}
where each term is given by 
\begin{eqnarray}
{\cal L}_{lepton}  & =& \int d^{4}\theta\;\left[  \,
\hat{\bar{L}}_{iL}e^{2g\hat{V}+g^{\prime}\left(  -\frac{1}{2}\right)  \hat{V}^{\prime}}\hat{L}_{iL}+
\hat{\bar{l}}_{iR}e^{g^{\prime}\left(  \frac{2}{2}\right) \hat{V}^{\prime}}\hat{l}_{iR}+ 
\hat{\bar{N}}_{iR}e^{g^{\prime}\left(  \frac{0}{2}\right) \hat{V}^{\prime}}\hat{N}_{iR} 
\,\right]
\,\ ,\nonumber \\
{\cal L}_{Quarks}  & =& \int d^{4}\theta\;\left[  \,
\hat{\bar{Q}}_{iL}e^{2g_{s}\hat{V}_{c}+2g\hat{V}+g^{\prime}\left(  \frac{1}{6}\right)\hat{V}^{\prime}}\hat{Q}_{iL}+
\hat{\bar{u}}_{iR}e^{2g_{s}\hat{V}_{c}+g^{\prime}\left(  -\frac{2}{3}\right)  \hat{V}^{\prime}}\hat{u}_{iR}\right.
\nonumber\\
& +&\left.  
\hat{\bar{d}}_{iR}e^{2g_{s}\hat{V}_{c}+g^{\prime}\left(\frac{1}{3}\right)  \hat{V}^{\prime}}\hat{d}_{iR}\,\right]  \,\ ,\nonumber\\
{\cal L}_{Gauge}  & =& \frac{1}{4} \left \{ \int d^{2}\theta\;\left[  
\sum_{a=1}^{8}W_{s}^{a\alpha}W_{s\alpha}^{a}+
\sum_{i=1}^{3}W^{i\alpha}W_{\alpha}^{i}+W^{\prime\alpha}W_{\alpha}^{\prime} \right]\,+h.c.\right\} \,, \nonumber \\
{\cal L}_{Higgs}  & =& \int d^{4}\theta\;\left[  \,\hat{\bar{H}}_{1}%
e^{2g\hat{V}+g^{\prime}\left(  -\frac{1}{2}\right)  \hat{V}^{\prime}}\hat
{H}_{1}+\hat{\bar{H}}_{2}e^{2g\hat{V}+g^{\prime}\left(  \frac{1}{2}\right)
\hat{V}^{\prime}}\hat{H}_{2}\right]  \nonumber\\
& +& \int d^{2}\theta \,\ W+\int d^{2}\bar{\theta} \,\ \bar{W}.
\label{supersymmetricpiece}
\end{eqnarray}
Therefore, our right handed neutrinos have any intercations with the usual gauge bosons then we will refere them as 
``fully sterile" \cite{Volkas:2001zb}. We use the notation $\hat{V}=T^{i}\hat{V}^{i}$ where $T^{i} \equiv \sigma^{i}/2$ (with $i=1,2,3$) are the generators of $SU(2)_{L}$ and 
$\hat{V}_{c} \equiv T^{a}\hat{V}^{a}_{c}$ and $T^{a}=\lambda^{a}/2$ (with $a=1,\cdots,8$) are the generators of $SU(3)_{C}$. As usual, 
$g_{s}$, $g$ and $g^{\prime}$ are the gauge couplings for the $SU(3)$, $SU(2)$ and $U(1)$ groups, 
and $W$ is the superpotential of this model and we will present it below. The field strength are given by \cite{wb}
\begin{eqnarray}
W_{s\alpha}^{a}  & =&-\frac{1}{8g_{s}}\,\bar{D}\bar{D}e^{-2g_{s}\hat{V}_{c}^{a}}D_{\alpha}e^{2g_{s}\hat{V}_{c}^{a}}\,\ \alpha=1,2
\,\ ,\nonumber\\
W_{\alpha}^{i}  & =&-\frac{1}{8g}\,\bar{D}\bar{D}e^{-2g\hat{V}^{i}}D_{\alpha}e^{2g\hat{V}^{i}}
\,\ ,\nonumber\\
W_{\alpha}^{\prime}  & =&-\frac{1}{4}\,DD\bar{D}_{\alpha}\hat{V}^{\prime}\,\ .
\label{fieldstrength}
\end{eqnarray}

The masses of charged gauge boson are the same as given at MSSM and it is given by
\begin{eqnarray}
M^{2}_{W}= \frac{g^{2}}{4}(v_{1}^{2}+v_{2}^{2})=
\frac{g^{2}v_{2}^{2}}{4}(1+\tan^{2} \beta)= 
\frac{g^{2}v^{2}_{2}}{4}\sec^{2} \beta \,\ ,
\label{wmass}
\end{eqnarray}
the new free parameter $\beta$ is defined in the following way
\begin{eqnarray}
\tan \beta \equiv \frac{v_{1}}{v_{2}},
\label{defbetapar}
\end{eqnarray}
where $v_{1}$ is the vev of $H_{1}$, this scalar give masses to quarks type $u$, while $v_{2}$ is the vev of the 
$H_{2}$, this scalar give masses to quarks type $d$ and for charged leptons. Due the fact that $v_{1}$ and $v_{2}$ are both positive, it imples that 
\begin{equation}
0 \leq \beta \leq \left( \frac{\pi}{2} \right) \,\ \mbox{rad}.
\label{constraintinbeta}
\end{equation}
Using the free parameter $\beta$, we can get the following relation to the vev o the Higgses defined at Eq.(\ref{vevhiggs})
\begin{eqnarray}
v_{1}&=& \frac{2M_{W}\sin \beta}{g}, \nonumber \\
v_{2}&=& \frac{2M_{W}\cos \beta}{g}.
\end{eqnarray}

In this case, the most general superpotential $W$ of this model is given by
\begin{eqnarray}
W_{MSSM3RHN}&=& \left[ W_{1N} + 
\left( \frac{W_{H}+W_{2RV}+W_{2N}}{2} \right) 
\right. \nonumber \\ &+& \left.
\left( \frac{W_{Y}+W_{3RV}+W_{3NRC}+W_{3NRV}}{3} \right) +hc \right], 
\label{mostgeneralsuperpotentialmssm3rhn}
\end{eqnarray}
where $W_{H},W_{Y}$ are defined at Eq.(\ref{Y-H}), while $W_{2RV},W_{3RV}$ from Eq.(\ref{mssmrpv}) and we have defined
\begin{eqnarray}
W_{1N}&=&\sum_{i=1}^{3}\kappa^{\prime}_{i}\hat{N}_{iR}, \nonumber \\
W_{2N}&=&\sum_{i,j=1}^{3}M^{\nu}_{ij} \hat{N}_{iR} \hat{N}_{jR}, \nonumber \\
W_{3NRC}&=&\sum_{i,j=1}^{3} f^{\nu}_{ij}\left( \hat{H}_{2}\hat{L}_{iL} \right) \hat{N}_{jR}, \nonumber \\ 
W_{3NRV}&=&\sum_{i,j,k=1}^{3} \left[ \kappa^{\prime \prime}_{i}\left( \hat{H}_{1}\hat{H}_{2} \right) \hat{N}_{iR}+ \kappa_{ijk} \hat{N}_{iR}  \hat{N}_{jR}\hat{N}_{kR} \right] . 
\end{eqnarray}
We can conclude the masses of charged fermions are the same as in the MSSM, and we can write \cite{Baer:2006rs,dress}
\begin{eqnarray}
f^{d}_{ij}&=&\frac{g}{\sqrt{2}M_{W}\cos \beta}m^{d}_{ij}, \,\
f^{l}_{ij}=\frac{g}{\sqrt{2}M_{W}\cos \beta}m^{l}_{ij}, \nonumber \\
f^{u}_{ij}&=&\frac{g}{\sqrt{2}M_{W}\sin \beta}m^{u}_{ij}.
\label{Yukawacouplingchargedleptonmasses}
\end{eqnarray}

The fact that $m_{u},m_{d},m_{s}$ and $m_{e}$ are many orders of magnitude 
smaller than the masses of others fermions may well be indicative of a 
radiative mechanism at work for these masses as considered at 
\cite{banks,ma}. We can explain the mass hierarchy in the charged 
fermion masses as showed in \cite{cmmc,cmmc1}.

We defined at Tab.(\ref{allrpqchargesinMSSM3RHN}) the $R$-charges of the superfields in the MSSM3RHN.  
\begin{table}[h]
\begin{center}
\begin{tabular}{|c|c|c|}
\hline  
$\mbox{ Superfield}$ & $R$-charge & $(B-L)$-charge \\
\hline
$\hat{L}_{iL}=( \hat{\nu}_{iL}, \hat{l}_{iL})^{T}$ & $n_{L}=+ \left( \frac{1}{2} \right)$ & $- \left( 1 \right)$ \\ 
\hline
$\hat{Q}_{iL}=(\hat{u}_{iL}, \hat{d}_{iL})^{T}$ & $n_{Q}=+ \left( \frac{1}{2} \right)$ & $+ \left( \frac{1}{3} \right)$ \\ 
\hline
$\hat{H}_{1}=( \hat{H}_{1}^{0},\, \hat{H}_{1}^{-})^{T}$ & $n_{H_{1}}=0$ & $0$  \\ 
\hline
$\hat{H}_{2}=( \hat{H}^{+}_{2}, \hat{H}^{0}_{2})^{T}$ & $n_{H_{2}}= 0$ & $0$ \\ 
\hline
$\hat{l}_{iR}$ & $n_{l}=- \left( \frac{1}{2} \right)$ & $+ \left( 1 \right)$  \\
\hline
$\hat{N}_{iR}$ & $n_{N}=- \left( \frac{1}{2} \right)$ & $+ \left( 1 \right)$  \\ 
 \hline
$\hat{d}_{iR}$ & $n_{d}=- \left( \frac{1}{2} \right)$ & $- \left( \frac{1}{3} \right)$  \\
\hline
$\hat{u}_{iR}$ & $n_{u}=- \left( \frac{1}{2} \right)$ & $- \left( \frac{1}{3} \right)$ \\ 
\hline
\end{tabular}
\end{center}
\caption{\small $R$-charge and $(B-L)$-charge assignment to all superfields in the MSSM3RHN.}
\label{allrpqchargesinMSSM3RHN}
\end{table}
In this model the right-handed sneutrinos are viable candidate to Dark Matter was showed at \cite{Pagetese,Abel:2006hr}

Using the $R$-charges defined at Tab.(\ref{allrpqchargesinMSSM3RHN}), will forbid the Majorana Mass term as well the 
$\Xi_{i},\lambda_{i},\kappa_{ijk},\mu_{i},\lambda_{ijk},\lambda^{\prime}_{ijk},\lambda^{\prime\prime}_{ijk}$ couplings 
in the superpotential defined at Eq.(\ref{mostgeneralsuperpotentialmssm3rhn}). In this case our superpotential become
\begin{eqnarray}
W&=&\left( \frac{W_{H}}{2} \right) +
\left( \frac{W_{Y}+W_{3NRC}}{3} \right) +hc,
\label{superpotentialmssm3rhnRPC}
\end{eqnarray}
see Eq.(\ref{mostgeneralsuperpotentialmssm3rhn}) and we can now write, in similar way a we have done at Eq.(\ref{Yukawacouplingchargedleptonmasses})
\begin{eqnarray}
f^{\nu}_{ij}&=&\frac{g}{\sqrt{2}M_{W}\sin \beta}m^{\nu}_{ij}.
\label{Yukawacouplingneutrinosmasses}
\end{eqnarray}
It is a nice result because now neutrinos are Dirac fermions \cite{Pagetese} and it was shown recently that the 
right-handed sneutrinos can be non-thermal in the presence of a Majorana mass term, as showed at \cite{degouvea}.

From Eqs.(\ref{superpotentialmssm3rhnRPC},\ref{Yukawacouplingneutrinosmasses}), we can write the simple relation between Yukawa coupling 
to neutrinos and their physical masses
\begin{equation}
m^{\nu}=f^{\nu}v,
\end{equation}
where $v \sim {\cal O}(10^{11})$eV is the vev of the electroweak breaking scale and if we impose 
\begin{equation}
m^{\nu}\sim {\cal O}(10^{-2}-10^{-1})
\end{equation} 
it imposes that the Yukawa coupling of neutrinos should be
\begin{equation}
f^{\nu}\sim {\cal O}(10^{-12}-10^{-13}),
\end{equation}
and in this case the non-thermal sterile sneutrino can be a good dark matter candidate see \cite{Pagetese,Allahverdi:2007wt,Page:2007sh}.

The experimental evidence suggests that the supersymmetry is not an exact symmetry. Therefore, supersymmetry 
breaking terms should be added to the Lagrangian defined by the Eq.(\ref{totallagmssm3rhn}). The most general soft supersymmetry 
breaking terms, which do not induce quadratic divergence, where described by Girardello and Grisaru \cite{10}. They 
found that the allowed terms can be categorized as follows: a scalar field $A$ with mass terms
\begin{equation}
{\cal L}^{SMT}=- A^{\dagger}_{i}m^{2}_{ij}A_{j},
\end{equation}
 a fermion field gaugino  $\lambda$ with mass  terms
\begin{equation}
{\cal L}^{GMT}=- \frac{1}{2} (M_{ \lambda} \lambda^{a} \lambda^{a}+hc),
\end{equation}
and finally trilinear scalar interaction terms
\begin{equation}
{\cal L}^{INT}= B_{ij}\mu_{ij}A_{i}A_{j}+A_{ijk}f_{ijk}A_{i}A_{j}A_{k}+hc \,\ .
\end{equation}
The terms in this case are similar with the terms allowed in the superpotential of the model we are going to consider next. As 
we are not doing any phenomenology with the strong sector, we are ommiting the squarks contribution because they are the same as in the MSSM.

The general soft SUSY breaking terms are given as
\begin{eqnarray}
{\cal L}^{SMT}_{MSSM3RHN}&=&- \left(
\sum_{i=1}^{3} \left[ 
\tilde{L}^{\dagger}_{iL} \left( M^{2}_{L}\right)_{ij}\tilde{L}_{jL}+
\tilde{l}^{\dagger}_{iR} \left( M^{2}_{l}\right)_{ij}\tilde{l}_{jR} +
\tilde{\nu}^{\dagger}_{iR}\left( M^{2}_{N} \right)_{ij} \tilde{\nu}_{jR} \right] 
\right. \nonumber \\ &+& \left.   
M^{2}_{H_{1}}|H_{1}|^{2}+  M^{2}_{H_{2}}|H_{2}|^{2} \right),
\end{eqnarray}
we do not write the masses to squarks. We can suppose there is just one soft SUSY breaking slepton mass for each generation of left-slepton, it means
\begin{equation}
\left( M^{2}_{L}\right)_{11}=M^{2}_{\tilde{e}_{L}}=M^{2}_{\tilde{\nu}_{eL}},
\left( M^{2}_{l}\right)_{11}=M^{2}_{\tilde{e}_{R}}, 
\left(M^{2}_{N}\right)_{11}=M^{2}_{\tilde{\nu}_{eR}}. 
\end{equation}
the terms similar to our superpotential, defined at Eq.(\ref{superpotentialmssm3rhnRPC}), is written as
\begin{eqnarray}
{\cal L}^{INT}_{MSSM3RHN}&=&
B \mu H_{1} H_{2}+ 
A^{l}_{ij}f^{l}_{ij} ( \tilde{L}_{iL}H_{1}) \tilde{l}_{jR}
+A^{\nu}_{ij}f^{\nu}_{ij} ( \tilde{L}_{iL}H_{2}) \tilde{\nu}_{jR}+hc, 
\label{softtermsmssm3rhn}
\end{eqnarray}
again we do not write the terms to squarks. The term ${\cal L}^{GMT}$ is the same as given at MSSM 
due it we do not presented it here and there expansion can be found at \cite{Baer:2006rs,dress}.

The only interaction to the right-handed sneutrinos are givn by is given by $( \tilde{L}_{iL}H_{2}) \tilde{\nu}_{jR}$ and the last 
term at Eq.(\ref{softtermsmssm3rhn}) and due this fact we can call the right-handed neutrinos as {\it sterile} snautrinos, see comment 
after Eq.(\ref{supersymmetricpiece}), as done at \cite{Page:2007sh}. 

\section{Masses}

We will here consider the masses to neutrinos and sneutrinos. The masses to charged leptons and charged sleptons are the same as in the 
MSSM, and can be found at \cite{Rodriguez:2019mwf,Baer:2006rs,dress}, due it we do not discuss them in this review.

\section{Masses to Neutrinos}
\label{sec:neutrinomasses}

In general case, we have three left-handed neutrinos and three right-handed neutrinos and taken into account our superpotential, see 
Eq.(\ref{superpotentialmssm3rhnRPC}), the most general terms to neutrinos are given by:
\begin{eqnarray}
- \left[ \frac{1}{3} f^{\nu}_{ij} \left( L_{iL}H_{2}\right) \nu_{jR}+ \frac{1}{2}\left( M_{N}\right)_{ij}\nu_{iR}\nu_{jR} \right] +hc,
\label{neutrinosmasses}
\end{eqnarray}
The parameters $f^{\nu}_{ij}$ are symmetric in $i,j$ exchange and we have choosen a basis for the right-handed neutrinos superfields 
so that the superpotential terms are diagonal, and $M \sim M_{Planck}$ or comparable to the SO(10) breaking scale as presented at \cite{Baer:2006rs}. 
After the electroweak symmetry is broken we get 
$6 \times 6$ mass matrix to diagonalize. This matrix can be write as
\begin{equation}
\left( 
\begin{array}{cc}
0 & M^{Dirac}_{\nu} \\
(M^{Dirac}_{\nu})^{T} & M
\end{array}
\right),
\label{treelevelmassmatrixtoneutrinos}
\end{equation}
where
\begin{eqnarray}
\left( M^{Dirac}_{\nu} \right)_{ij}&=&\frac{1}{3} f^{\nu}_{ij} v_{2}, \nonumber \\
\left( M \right)_{ij}&=&\frac{1}{2} \left( M_{N}\right)_{ij}.
\label{diracmasstermstoneutrinos}
\end{eqnarray}
We get the usual see saw mechanism (type I) \cite{sees1} if $M^{Dirac}_{\nu} \ll M$, the mass spectrum is one heavy Majorana neutrino, with mass $M$. We 
get three light Dirac neutrinos with their masses given by
\begin{equation}
m_{\nu} \simeq \frac{(M^{Dirac}_{\nu})^{2}}{M}.
\label{neutrinosmassesattreelevel}
\end{equation}
and the neutrinos are Majorana particles as happen in MSSM.   

One interesting case, happen when we consider three left-handed neutrinos and only one right-handed neutrinos. Under this consideration we get the 
mass matrix from Eq.(\ref{neutrinosmasses})
\begin{eqnarray}
- \left[ \frac{v_{2}}{3 \sqrt{2}} \left( f^{\nu}_{1} \nu_{1L} \nu_{R}+ f^{\nu}_{2} \nu_{2L} \nu_{R}+ f^{\nu}_{3} \nu_{3L} \nu_{R} \right) 
+ \frac{1}{2}M_{N}\nu_{R}\nu_{R} \right] +hc,
\label{neutrinosmassescontribution}
\end{eqnarray}
in the base $( \nu_{1L}, \nu_{2L}, \nu_{3L},\nu_{R})^{T}$ we get the mass matrix presented at Eq.(\ref{treelevelmassmatrixtoneutrinos}), and in 
this case we can write \cite{King:1998jw,Davidson:1998bi}
\begin{equation}
M^{Dirac}_{\nu}= \frac{v_{2}}{3 \sqrt{2}} \left( 
\begin{array}{ccc}
(f^{\nu}_{1})^{2} & f^{\nu}_{1}f^{\nu}_{2} & f^{\nu}_{1}f^{\nu}_{3} \\
f^{\nu}_{1}f^{\nu}_{2} & (f^{\nu}_{2})^{2} & f^{\nu}_{2}f^{\nu}_{3} \\
f^{\nu}_{1}f^{\nu}_{3} & f^{\nu}_{2}f^{\nu}_{3} & (f^{\nu}_{3})^{2}
\end{array}
\right),
\label{neutrino left-rightMSSMRN}
\end{equation}
we can draw at this case a diagramatic representation similar to ones shown at Fig.(\ref{rpvt}). We make the additional hipotesis, all the Yukawa 
coupling $f^{\nu}_{i}$ are real. From Eq.(\ref{neutrino left-rightMSSMRN}) it is simple to show the following results
\begin{eqnarray}
\det \left( M^{Dirac}_{\nu} \right)&=&0, \nonumber \\
\mbox{Tr}\left( M^{Dirac}_{\nu} \right)&=&(f^{\nu}_{1})^{2}+(f^{\nu}_{2})^{2}+(f^{\nu}_{3})^{2}.
\end{eqnarray}
The characteristic equation obtained from Eq.(\ref{neutrino left-rightMSSMRN}) is
\begin{equation}
\det \left( M^{Dirac}_{\nu}- \lambda I_{3 \times 3} \right)=- \lambda^{3}+ \lambda^{2}\left[ (f^{\nu}_{1})^{2}+(f^{\nu}_{2})^{2}+(f^{\nu}_{3})^{2} \right] =0,
\end{equation}
it means we have one more zero eigenvalue. Therefore, we can conclude, Eq.(\ref{neutrino left-rightMSSMRN}) has two 
zero eigenvalues and one non-zero eigenvalue, as presented at \cite{King:1998jw,Davidson:1998bi}.

The mass eigenstate, to a single Dirac neutrino represented as $\nu_{L}$, is defined in the following way \cite{King:1998jw,Davidson:1998bi}
\begin{equation}
\nu_{H}= \frac{f^{\nu}_{1}\nu_{1L}+f^{\nu}_{2}\nu_{2L}+f^{\nu}_{3}\nu_{3L}}{\sqrt{(f^{\nu}_{1})^{2}+(f^{\nu}_{2})^{2}+(f^{\nu}_{3})^{2}}},
\label{diracneutrinoeigenvector}
\end{equation}
and their masses is
\begin{equation}
m_{\nu_{H}}=\left[ (f^{\nu}_{1})^{2}+(f^{\nu}_{2})^{2}+(f^{\nu}_{3})^{2} \right]
\frac{v_{2}}{3 \sqrt{2M}},
\label{diracneutrinoeigenvalue}
\end{equation}
where $M$ is the Majorana mass of our singlet $\nu_{R}$. It is the 
seesaw mechanism (type I) \cite{sees1}  to give mass to neutrinos, alternatives to the seesaw mechanism was presented at 
\cite{Smirnov:2004hs}. 

We can, also, with this simple mechanism explain the atmospheric neutrino 
data in the following way, our completely free parameters, $f^{\nu}_{i}$, 
must satisfy the following relation \cite{King:1998jw}
\begin{eqnarray}
f^{\nu}_{1}\ll f^{\nu}_{2}\approx f^{\nu}_{3},
\end{eqnarray}
we can define a small angle $\theta_{1}$ and a large angle 
$\theta_{23}\equiv \theta_{atm}$
\footnote{See Introduction at Eq.(\ref{eq1:best-fit_mass}) about the second angle.} as:
\begin{equation}
\nu_{H}= \theta_{1}\nu_{eL}+ \sin \theta_{23}\nu_{\mu L}+
\cos \theta_{23}\nu_{\tau L},
\label{atmexplanation}
\end{equation}
where using Eq.(\ref{diracneutrinoeigenvector}) allows to redifine
\begin{eqnarray}
\theta_{1}\approx 
\frac{f^{\nu}_{1}}{\sqrt{(f^{\nu}_{1})^{2}+(f^{\nu}_{2})^{2}+(f^{\nu}_{3})^{2}}}, \nonumber \\
\sin \theta_{23}\approx 
\frac{f^{\nu}_{2}}{\sqrt{(f^{\nu}_{1})^{2}+(f^{\nu}_{2})^{2}+(f^{\nu}_{3})^{2}}}, \nonumber \\
\cos \theta_{23}\approx 
\frac{f^{\nu}_{3}}{\sqrt{(f^{\nu}_{1})^{2}+(f^{\nu}_{2})^{2}+(f^{\nu}_{3})^{2}}},
\label{mixingparam}
\end{eqnarray}

We have done this approach because we already know to the Standard Model 
may account to explain the atmospheric neutrino data via 
$\nu_{\mu}\rightarrow \nu_{\tau}$ oscillations with the parameters 
defined at second line in Eq.(\ref{eq1:best-fit_mass}). Our conclusion 
to explain the atmospheric neutrino mixing is, we have a single massive 
neutrino $\nu_{H}$, defined at Eq.(\ref{atmexplanation}), with a 
Majorana mass, defined at Eq.(\ref{diracneutrinoeigenvalue}), and we will 
impose its numerical valus is given by $m_{\nu}\sim 5\times 10^{-2}\ eV$, 
and the mixing, defined at Eqs.(\ref{mixingparam}), is 
choose to be given by $\theta_{23} \sim \pi/4$ in agreement with the 
experimental data given at Eq.(\ref{tbm}).

On this scenario the two neutrino eigenstates which are orthogonal to the massive neutrino $\nu_{H}$ will be massless. We can also set, by 
convenienc, $\theta_1 =0$ in such way that we can choose as one of 
the massless neutrinos will be $\nu_{eL}$ and the other will be the
orthogonal combination 
\begin{eqnarray}
\nu_{0L}\approx \cos \theta_{23}{\nu}_{\mu L}- 
\sin \theta_{23}{\nu}_{\tau L}.
\label{nu0}
\end{eqnarray}

Now we are going to explaing the Pontecorvo-Maki-Nakagawa-Sakata (PMNS) matrix, $U_{PMNS}$, is defined at Eq.(\ref{pmnsform}), in our case is:
\begin{eqnarray}
\left(
\begin{array}{l}
\nu_{e}  \\
\nu_{\mu}  \\
\nu_{\tau} 
\end{array}
\right)_{L} = U 
\left(
\begin{array}{l}
\nu_{e}^{\prime} \\
\nu_{0}^{\prime} \\
\nu_{H}
\end{array}
\right)_{L}
\label{defns}
\end{eqnarray}
where $\nu_{H}$ is defined at Eq.(\ref{atmexplanation}). The matrix 
$U$ can be defined as \cite{King:1998jw}
\begin{eqnarray}
U = \left( \begin{array}{ccc}
 1 &  
 \left( \frac{\cos \theta_{23}}{\sin \theta_{23}}\right) \theta_{1}   
 & \theta_{1}      \\
- \left( \frac{\theta_{1}}{\sin \theta_{23}} \right)  & 
\cos \theta_{23} &  \sin \theta_{23}   \\
0   &  - \sin \theta_{23}   & \cos \theta_{23} 
\end{array}
\right)
\label{U}
\end{eqnarray}
In this basis the mass eigen-states expressed in terms of the weak
eigenstates are summarised below:
\begin{eqnarray}
\nu_{eL}^{\prime}&=& \nu_{eL} - 
\left( \frac{\theta_{1}}{\sin \theta_{23}} \right) \nu_{\mu L} 
\nonumber \\
\nu_{0 L}^{\prime}& = & \left( 
\frac{\cos \theta_{23}}{\sin \theta_{23}}\right) \theta_1 \nu_{eL} 
+ \cos \theta_{23} \nu_{\mu L} - \sin \theta_{23}\nu_{\tau L} 
\nonumber \\
\nu_{H} &=& \theta_{1}{\nu}_{eL} 
+ \sin \theta_{23}{\nu}_{\mu L}
+ \cos \theta_{23}{\nu}_{\tau L}
\nonumber \\
\end{eqnarray}
In the $\theta_1=0$ limit $\nu_{eL}^{\prime},\nu_{0 L}^{\prime}$
return to $\nu_{eL},\nu_{0 L}$ defined at Eq.(\ref{nu0}). Details about 
numerical analyses see \cite{King:1998jw}.

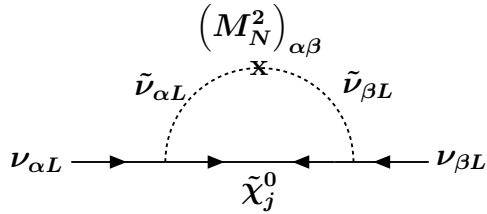
\begin{figure}[ht]
\unitlength.5mm
\SetScale{1.418}
\begin{boldmath}
\begin{center}
\begin{picture}(120,40)(0,0)
\ArrowLine(0,0)(25,0)
\ArrowLine(25,0)(52,0)
\ArrowLine(70,0)(52,0)
\ArrowLine(95,0)(70,0)
\DashCArc(50,0)(25,0,180){1}
\Text(-2,0)[r]{$\nu_{\alpha L}$}
\Text(97,0)[l]{$\nu_{\beta L}$}
\Text(50,25)[c]{{\bf x}}
\Text(24,20)[c]{$\tilde{\nu}_{\alpha L}$}
\Text(79,20)[c]{$\tilde{\nu}_{\beta L}$}
\Text(50,35)[c]{$\left(M^{2}_{N}\right)_{\alpha \beta}$}
\Text(50,-8)[c]{$\tilde{\chi}^{0}_{j}$}
\end{picture}
\end{center}
\end{boldmath}
\vspace{10mm}
\caption{One-loop diagram generating  a Majorana neutrino mass adapted from \cite{Davidson:1998bi,GH}.}
\label{f4}
\end{figure}

The others two mass neutrinos $\nu_{eL}, \nu_{0L}$ get their masses from 1-loop level. 
Having generated a sneutrino Majorana mass, 
see Sec.(\ref{sec:sneutrinosmasses}), it is straightforward to
see that such a mass will lead to 1-loop radiative corrections to 
neutrino masses via the diagram drawn in Fig.(\ref{f4}) \cite{GH}.
The mass expression to our neutrinos can be written as \cite{GH}
\begin{eqnarray} 
m_\nu^{(1)} = 
\frac{g^{2} m_{\tilde{\nu}}}{32 \pi^{2} \cos^{2} \theta_{W}}
\sum_{j} \,f(y_{j}) |Z_{jZ}|^{2}\,,
\label{1loopneutrinosmasses}
\end{eqnarray}
where $m_{\tilde{\nu}}$ is the sneutrinos masses and we have defined
\begin{eqnarray} 
f(y_{j}) = 
\frac{\sqrt{y_{j}}
\left[y_{j}-1-\ln(y_{j})\right]}{(1-y_{j})^{2}},
\label{loopmass}
\end{eqnarray}
where 
\begin{eqnarray} 
y_{j} \equiv \frac{m_{\tilde{\nu}}^{2}}{m_{\tilde\chi^0_j}^{2}},
\end{eqnarray}
and
$Z_{jZ}\equiv Z_{j2}\cos\theta_{W}-Z_{j1}\sin\theta_{W}$ is the
neutralino mixing matrix element \cite{GH}. 

We can with this mechanism explain neutrinos data as shown at \cite{King:1998jw,Davidson:1998bi}.

\section{Masses of Sneutrinos}
\label{sec:sneutrinosmasses}

The masses and mixing of sparticles are of crucial importance both experimentally and theoretically \cite{Rodriguez:2019mwf,Baer:2006rs,dress}. 
All the terms that give contribution to sneutrino mass matrix came from Eq.(\ref{softtermsmssm3rhn}), it means that
the mixing in this sector is unrelated with the mixing we get at neutrino sector. In general on this sector we get one 
$6 \times 6$ mass matrix (we get Super-CKM and Super-PMNS matrix \cite{dress,9604378}), but when we ingone the mixing into 
families this matrix can be devided into three blocks of 
$2 \times 2$ for each family, we will use this fact here, as discussed at \cite{Baer:2006rs,dress}.

As we have done above, we will consider a one generation model. The 
relevant superpotential is defined, we are going present a review from 
the main results presented at \cite{GH} but using my notation, is again 
defined at Eq.(\ref{neutrinosmasses}) and the relevant terms in the soft 
supersymmetry breaking are given by
\begin{eqnarray}
V_{soft}=\left[ m^{2}_{\tilde{\nu}_{eL}}\tilde{\nu}^{\dagger}_{eL}\tilde{\nu}_{eL}+
m^{2}_{\tilde{\nu}_{eR}}\tilde{\nu}^{\dagger}_{eR} \tilde{\nu}_{eR}
+\left( 
A^{\nu}f^{\nu}H_{2} \tilde{\nu}_{eL} \tilde{\nu}_{eR}+
MB \tilde{\nu}_{eR} \tilde{\nu}_{eR}+hc \right) \right]. \nonumber \\
\label{masssneutrinosmajoranamassterm}
\end{eqnarray}

The sneutrino masses are obtained by diagonalizing a 
$4\times 4$ squared-mass matrix. In similar way as we perform when we 
diagonalize the Higgs mass matrix at MSSM, it is convenient to define: 
\begin{eqnarray}
\tilde{\nu}&=& \frac{ \left(
\tilde{\nu}_{1}+ \imath \tilde{\nu}_{2} \right)}{\sqrt{2}}, \,\
\tilde{N}= \frac{ \left(
\tilde{N}_{1}+ \imath \tilde{N}_{2}\right)}{\sqrt{2}}.
\end{eqnarray} 
Then, the squared-sneutrino mass matrix (${\cal M}^{2}$), in similar way 
as done in the Higgs mass matrix, can be separates into CP-even and 
CP-odd blocks:
\begin{eqnarray}
{\cal M}^{2}=\frac{1}{2}
\left( 
\begin{array}{cc} 
\phi_{1} & \phi_{2} 
\end{array}\right)
\left( 
\begin{array}{cc} 
{\cal M}^2_{+} & 0 \\
0 & {\cal M}^2_{-}
\end{array}\right)
\left( 
\begin{array}{c} 
\phi_{1} \\ 
\phi_{2} 
\end{array}\right)
\,,
\end{eqnarray}
where we have using the following simple notation 
\begin{eqnarray}
\phi_{i}\equiv \left( 
\begin{array}{cc} 
\tilde{\nu}_{i} & \tilde{N}_{i}
\end{array}\right),
\end{eqnarray}
and
\begin{eqnarray}
{\cal M}^2_{\pm} = \left( 
\begin{array}{cc}
m_{\tilde L}^2 + \frac{1}{2} m_Z^2 \cos 2\beta + m_D^2 &
m_D[A_\nu-\mu \cot\beta\pm M] \\
m_D[A_\nu-\mu \cot\beta\pm M] & M^2 + m_D^2 + m_{\tilde N}^2 \pm 2 BM  
\end{array}\right),
\end{eqnarray}
where $M$ is the largest mass parameter and $m_D$ is the Dirac mass term 
to neutrinos defined at Eq.(\ref{diracmasstermstoneutrinos}). 
Then, to first order in $(1/M)$,
the two light sneutrino eigen-states are $\tilde{\nu}_{1}$ and 
$\tilde{\nu}_{2}$, with corresponding squared masses:
\begin{eqnarray}
m^{2}_{\tilde{\nu}_{1,2}} = m_{\tilde L}^{2}+
\frac{1}{2} m_{Z}^{2} \cos {2}\beta\mp \frac{1}{2}\Delta m^{2}_{\tilde{\nu}}\,,
\end{eqnarray}
where the squared mass difference $\Delta m^{2}_{\tilde{\nu}}\equiv
m^{2}_{\tilde{\nu}_{2}}-m^{2}_{\tilde{\nu}_{1}}$ is of order $(1/M)$.

The mass term to sneutrinos, if we do not allow Majorana Mass terms to 
right-handed neutrinos is given by
\begin{eqnarray}
{\cal L}^{sneutrinos}_{masses}&=&- \left[
m^{2}_{\tilde{\nu}_{eL}}\tilde{\nu}^{\dagger}_{eL}\tilde{\nu}_{eL}+
m^{2}_{\tilde{\nu}_{eR}}\tilde{\nu}^{\dagger}_{eR} \tilde{\nu}_{eR}
+A^{\nu}f^{\nu}v_{2} \left( \tilde{\nu}_{eL} \tilde{\nu}_{eR}+hc \right) \right]. \nonumber \\
\label{sneutrinomixing}
\end{eqnarray}
We have defined
\begin{eqnarray}
m^{2}_{\tilde{\nu}_{eL}}&=&M^{2}_{L}+ \frac{1}{2}\cos \left( 2 \beta \right) M^{2}_{Z}, \nonumber \\
m^{2}_{\tilde{\nu}_{eR}}&=&M^{2}_{N},
\end{eqnarray}
where $M_{Z}$, as usual, is the $Z$ boson mass. The first line above is similar as done in the MSSM when we calculate the sneutrinos 
masses \cite{Rodriguez:2019mwf,Baer:2006rs,dress}. More general case, with Majorana mass term to neutrinos is dicussed at 
\cite{GH}.We must to emphasize that the paraqmeters $M^{2}_{L},M^{2}_{N}$ and $A^{\nu}$ are defined at 
electroweak scale because they break softly the SUSY.

In order to get the sneutrinos mass eigenstates, we have, first, to diagonalize the following mass matrix
\begin{equation}
M^{2}_{sneutrinos}= \left(
\begin{array}{cc}  
m^{2}_{\tilde{\nu}_{eL}}  & A^{\nu}f^{\nu}v_{2} \\
A^{\nu}f^{\nu}v_{2} & m^{2}_{\tilde{\nu}_{eR}} 
\end{array}\right),
\end{equation}
it is simple to show the following results
\begin{eqnarray}
\det \left( M^{2}_{sneutrinos} \right)&=&m^{2}_{\tilde{\nu}_{eL}}m^{2}_{\tilde{\nu}_{eR}}- \left( A^{\nu}f^{\nu}v_{2} \right)^{2}, \nonumber \\
\mbox{Tr}\left( M^{2}_{sneutrinos} \right)&=&m^{2}_{\tilde{\nu}_{eL}}+m^{2}_{\tilde{\nu}_{eR}}.
\end{eqnarray}
The characteristic equation obtained from Eq.(\ref{neutrino left-rightMSSMRN}) is
\begin{equation}
\det \left( M^{2}_{sneutrinos}- \lambda I_{2 \times 2} \right)= \lambda^{2}- \left( m^{2}_{\tilde{\nu}_{eL}}+m^{2}_{\tilde{\nu}_{eR}} \right) \lambda 
+m^{2}_{\tilde{\nu}_{eL}}m^{2}_{\tilde{\nu}_{eR}}- \left( A^{\nu}f^{\nu}v_{2} \right)^{2} =0.
\end{equation}
it means we have two massives eigenstates.

The eigenvalues of $M^{2}_{sneutrinos}$ are given by
\begin{equation}
m^{2}_{\pm}=M^{2}\pm \sqrt{\left( m^{2} \right)^{2}+ \left( A^{\nu}f^{\nu}v_{2} \right)^{2}},
\end{equation}
where we have defined the following parameters
\begin{eqnarray}
M^{2}&=& \left( \frac{m^{2}_{\tilde{\nu}_{eL}}+m^{2}_{\tilde{\nu}_{eR}}}{2} \right), \nonumber \\
m^{2}&=& \left( \frac{m^{2}_{\tilde{\nu}_{eL}}-m^{2}_{\tilde{\nu}_{eR}}}{2} \right).
\end{eqnarray}

The eigenvectors are given by \cite{Pagetese}
\begin{eqnarray}
\left( 
\begin{array}{c} 
\tilde{\nu}_{e+} \\ 
\tilde{\nu}_{e-} 
\end{array}\right) =
\left( 
\begin{array}{cc} 
\cos{\theta_{\tilde{\nu}}} & \sin{\theta_{\tilde{\nu}}}  \\ 
- \sin{\theta_{\tilde{\nu}}} & \cos{\theta_{\tilde{\nu}}} 
\end{array}\right)
\left( 
\begin{array}{c} 
\tilde{\nu}_{eL} \\ 
\tilde{\nu}^{\dagger}_{eR} 
\end{array}\right) , \nonumber
\label{physicalsneutrinos}
\end{eqnarray}
where we define the mixing angle as
\begin{eqnarray}
\cos{\theta_{\tilde{\nu}}}&=&
\frac{m^{2}+\sqrt{\left( m^{2} \right)^{2}+ \left( A^{\nu}f^{\nu}v_{2} \right)^{2}}}{|\tilde{\nu}|^{2}}
, \nonumber \\
\sin{\theta_{\tilde{\nu}}}&=&\frac{\left( A^{\nu}f^{\nu}v_{2} \right)}{|\tilde{\nu}|^{2}}
, \nonumber \\
|\tilde{\nu}|^{2}&=& 
\left( m^{2}+\sqrt{\left( m^{2} \right)^{2}+ \left( A^{\nu}f^{\nu}v_{2} \right)^{2}}\right)^{2}+ \left( A^{\nu}f^{\nu}v_{2} \right)^{2},
\end{eqnarray}
in this model the $\tilde{\nu}_{e-} $ is LSP \cite{Pagetese}.

\section{Flat direction of MSSM3RHN Model}
\label{sec:cosmological}

We can calculate all flat directions in the MSSM using the prescription given in  \cite{Gherghetta:1995dv}.
The flat directions $\hat{N}$ and $\hat{L}\hat{H}_{2}$ will generate a left-right asymmetry in the sneutrino sector \cite{Pagetese,Abel:2006hr}. 

The flat directions shown in Tab.(\ref{tab:flatdir}) was getting using the most general superpotential of this model presented at 
Eq.(\ref{mostgeneralsuperpotentialmssm3rhn}).

\begin{table}[htb]
\renewcommand{\arraystretch}{1.10}
\begin{center}
\normalsize
\vspace{0.5cm}
\begin{tabular}{|c|c|}
\hline
Flat direction & $(B-L)$ \\
\hline
$\hat{N}$ & 1 \\
\hline
$\hat{N}\hat{N}$ & 2 \\
\hline
$\hat{H}_{1}\hat{H}_{2}$ & 0 \\
\hline
$\hat{L}\hat{H}_{2}$ & -1 \\
\hline
$\hat{N}\hat{N}\hat{N}$ & 3 \\
\hline
$\hat{H}_{1}\hat{H}_{2}\hat{N}$ & 1 \\
\hline
$\hat{L}\hat{H}_{2}\hat{N}$ & 0 \\
\hline
\end{tabular}
\caption{\small Flat direction of the model MSSM3RHN.}
\label{tab:flatdir} 
\end{center}
\end{table}

\section{Supersymmetric $B-L$ Model (SUSYB-L)}

The TeV scale right-handed neutrino is naturally obtained in supersymmetric $B-L$ ($SUSYB-L$) extension of the SM and it is one of the 
simplest model beyond the SM that provides a viable and testable solution to the neutrino mass and this class of model can account for the experimental 
results of the light neutrino masses and their large mixing angles. 

At this moment we will consider the Baryon number (B) minus Lepton number (L) the $B-L$ number as a 
gauge symmetry and this symmetry is broken at TeV scale \cite{Khalil:2007dr,Khalil:2006yi}, remember that it is the scale to break SUSY in order to 
explain the hierarchy problem at MSSM. In this kind of model, as happen in the MSSM, the Higgs potential receives large radiative corrections
that induce spontaneous $B-L$ symmetry breaking at TeV scale, in analogy to the electroweak symmetry breaking in MSSM \cite{Khalil:2007dr}. 

The interesting fact in this kind of model is, there are new complex phases in the leptonic sector can generate lepton asymmetry, which is converted 
to baryon asymmetry \cite{Khalil:2006yi} and the relevant terms necessary came from the following soft terms that also break SUSY \cite{Khalil:2009tm,Kajiyama:2009ae}
\begin{equation}
{\cal L}_{\rm soft} = \frac{\tilde m_{n}^{2}}{2} \left( \tilde{n}^{c}\right)^{\dag}
\tilde{n}^{c}+ \frac{B_{n}^{2}}{2} \tilde{n}^{c} \tilde{n}^{c} + 
A_{n}Y_{n} \left( \epsilon \tilde{L}H_{2} \right) \tilde{n^{c}}  + h.c. 
\label{egipciobl}
\end{equation}
There are a mixing between the sneutrino $\tilde{n}^{c}$ and the anti-sneutrino $\tilde{n}^{c\dag}$. The CP violation phase in this mixing 
generates lepton asymmetry in the final states of the $\tilde{n}^{c}$-decay and it is converted to baryon asymmetry through the sphaleron process 
\cite{spha}. 

\section{Supersymmetric Left-Right Model (SUSYLR)}
\label{sec:susylr}

One of the main motivation to study this kind of model are to explain the observed lightness of neutrinos in a natural way and it can 
also solve the strong CP problem \cite{phenosusylr}. On this model the gauge symmetry is given by
\begin{equation}
SU(2)_{L} \otimes SU(2)_{R} \otimes U(1)_{B-L}.
\end{equation}
The lepton content of the model is a little different of the lepton content in the MSSM where the 
left-handed fermions are in doublet representation while the right-handed are singlets of 
$SU(2)_{L}$. Now both the leptons, left-handed and right-handed, are in doublets in their $SU(2)$ gauge group as we show below
\begin{eqnarray}
\hat{L}_{i}\sim({\bf2},{\bf1},-1), 
\,\
\hat{L}^{c}_{i} \sim({\bf1},{\bf2},1), 
\end{eqnarray}
in parentheses it appears the transformation properties under the respective 
$(SU(2)_{L},SU(2)_{R},U(1)_{B-L})$. 

On the literature there are two different SUSYLR models: the first one uses $SU(2)_{R}$ triplets (SUSYLRT) \cite{susylr} and the
second $SU(2)_{R}$ doublets (SUSYLRD) \cite{doublet}. Unfortunatelly in SUSYLRD the neutrinos are no massive due it we will not present it on 
this review.

\subsection{Triplet Model (SUSYLRT)}
\label{apend:susylrt}

The scalars fields of this model are
\begin{eqnarray}
\hat{\Delta}_{L}&\sim& \left({\bf3},{\bf1},2 \right) , \,\ 
\hat{\Delta}^{\prime}_{L} \sim \left({\bf3},{\bf1},-2 \right) , \nonumber \\
\hat{\delta}^{c}_{L}&\sim& \left({\bf1},{\bf3},-2 \right) , \,\
\hat{\delta}^{\prime c}_{L}\sim \left({\bf1},{\bf3},2 \right) , \nonumber \\
\hat{\Phi}&\sim& \left({\bf2},{\bf2},0 \right) , \,\
\hat{\Phi}^{\prime} \sim\left({\bf2},{\bf2},0 \right) . 
\end{eqnarray}
The most general superpotential $W$ is given by \cite{susylr}
\begin{eqnarray}
W  & =&M_{\Delta}Tr(\hat{\Delta}_{L}\hat{\Delta}_{L}^{\prime})+
M_{\delta^{c}}Tr(\hat{\delta}_{L}^{c}\hat{\delta}_{L}^{\prime c})+
\mu_{1}Tr \left[ \left( \imath\sigma_{2}\right) \hat{\Phi}\left( \imath\sigma_{2}\right) \hat{\Phi} \right] +
\mu_{2}Tr\left[ \left( \imath\sigma_{2}\right) \hat{\Phi}^{\prime}\left( \imath\sigma_{2}\right) \hat{\Phi}^{\prime} \right] \nonumber\\
&+&
\mu_{3}Tr \left[ \left( \imath\sigma_{2}\right) \hat{\Phi}\left( \imath\sigma_{2}\right) \hat{\Phi}^{\prime} \right] +
f_{ab}Tr \left[ \hat{L}_{a}\left( \imath\sigma_{2}\right) \hat{\Delta}_{L}\hat{L}_{b} \right]+
f_{ab}^{c}Tr \left[ \hat{L}_{a}^{c}\left( \imath\sigma_{2}\right) \hat{\delta}_{L}^{c}\hat{L}_{b}^{c} \right]
\nonumber\\
&+&
h_{ab}^{l}Tr \left[ \hat{L}_{a}\hat{\Phi}\left( \imath\sigma_{2}\right) \hat{L}_{b}^{c} \right] +
\tilde{h}_{ab}^{l}Tr \left[ \hat{L}_{a}\hat{\Phi}^{\prime}\left( \imath\sigma_{2}\right) \hat{L}_{b}^{c} \right] .
\label{suplr}
\end{eqnarray}
Where $h^{l},\tilde{h}^{l}$ are the Yukawa couplings for the leptons. This model can be embedded in a supersymmetric
grand unified theory as $SO(10)$ and we will discuss some Supersymmetric Grand Unification models below. 

The masses to neutrinos is given by \cite{cmmc}:
\begin{eqnarray}
M_{ab}^{\nu}  & =&\frac{1}{\sqrt{2}}\left[  k_{1}h_{ab}^{l}+k_{2}^{\prime
}\tilde{h}_{ab}^{l}\right]  (\nu_{a}\nu_{b}^{c}+hc)+
\frac{\upsilon_{R}}{\sqrt{2}}f_{ab}^{c}(\nu_{a}^{c}\nu_{b}^{c}+hc)\nonumber\\
& -&\frac{\upsilon_{L}}{\sqrt{2}}f_{ab}(\nu_{a}\nu_{b} +h.c.).
\label{lepmasssusylr}
\end{eqnarray}
This result is in agreement with the presented in \cite{mfrank}, if we take
$\upsilon_{L}=0$.

\section{Supersymmetric Grand Unification Theory (SUSYGUT)}

In the standard minimal SUSYGUT scenario, the theory possesses both the
supersymmetry and the unified gauge symmetry at the unification scale. In our opinion the main properties of a finite SUSY GUT are \cite{GUT}:
\begin{itemize}
\item  the number of generations is fixed by the requirement of finiteness,
\item  various realistic possibilities are given by $SU(5),SU(6),SO(10)$ and $E(6)$ gauge groups,
\end{itemize}
from the last point, it is simple to coinvince yourself that there are a huge class of this kind of model. 
We will present, very short, two supersymmetric grand unification models that give masses to neutrinos, they are $SU(5)_{RN}$ and $SO(10)_{M}$.

\subsection{$SU(5)$ grand unified model with right-handed neutrinos $SU(5)_{RN}$}

Nice review about some general consideration about $SU(5)$ SUSY model are given at \cite{Bagnaschi:2016afc,deBoer:1994he}. Unfortunatelly 
the following channel to proton decay $p \rightarrow K^{+} \nu^{c}$ is enough to exclude the minimal supersymmetric SU(5) model \cite{mura02}. 
If we want to continue in the minimal $SU(5)$ supersymmetric model an interesting challenging is present at \cite{Bajc:2013dea}. Same in the 
contest of the last reference, the neutrinos are not massive and we need to change the lepton sector to get one mechanism to generate masses 
to the neutrinos.

In the  $SU(5)$ grand unified model with right-handed neutrinos $SU(5)_{RN}$, as in the minimal $SU(5)$ model we introduce the charged 
lepton\footnote{$5^{*}$ representation of $SU(5)$ and the indices $i,j$, as usual, denote generations} at $\Phi _{i} \sim \bar{5}$. We need also to 
introduce the right handed neutrino as $\eta _{i}\sim 1$. The scalars in this model are given by $H\sim 5$ and $\overline{H}\sim \bar{5}$. 
The superpotential of this model is given by \cite{Causse:2002ph,Causse:2002mu}:
\begin{equation}
\label{superpot2}
W_{SU(5)_{RN}}= f^{n}_{ij}\eta _{i}\Phi_{jA}H^{A}+ \frac{1}{2}M_{ij}\eta _{i} \eta _{j}.
\end{equation}
where we wrote only the terms that give neutrinos masses and the indices $i,j$ are familly indices while the $A$ is 
$SU(5)$ indice and run from $1$ to $5$. Therefore, $M_{ij}$ is the Majorana mass matrix for right handed neutrinos while 
$f^{n}$ is the Dirac mass term to neutrinos and both genearete mass via see-saw Mechanism. 

\subsection{Minimal SO(10) Supersymmetric Model $SO(10)_{M}$}

About the $SO(10)$ supersymmetric models see the following nice reviews \cite{bdr,Raby:2003in}. As we are interested in sutdy the generation to neutrino masses we will condider the model presented at \cite{Goh:2003hf,Goh:2003nv}. The lepton is introduced at 
$\psi_{i} \sim 16$ representation of $SO(10)$ while the Higgs fields $H_{10}\sim 10$ and $\Delta \sim 126^{*}$ representation of $SO(10)$. 
Their superpotential, responsable to give masses to usual lepton, is
\begin{eqnarray}
W_{SO(10)_{M}}&=&  h_{ij}\psi_{i}\psi_{j} H_{10} + f_{ij} \psi_{i}\psi_{j}\Delta
\end{eqnarray}
$h$ and $f$ are symmetric matrices in the indeces $i,j$ and, as usual, they denote generations indices. The neutrino masses in this model is given by
\begin{eqnarray}
M_{\nu^D} &=& \left( h^{*} \cos \alpha_{u}-3 f^{*} e^{\imath \gamma_{u}} \sin \alpha_{u} \right) \sin \beta , 
\label{sumrule}
\end{eqnarray}
where $\alpha_{u}$ is a new parameter in this model responsible for nonzero CKM mixing angles. For more details about the notation given at 
Eq.(\ref{sumrule}) see \cite{Goh:2003hf}.

When we consider $SO(10)$ Supersymmetric Model,  there are another intersting mechanism to generate neutrino masses using the scalar at 
$126$ representations of $SO(10)$ and about this mechanism we recommend to see \cite{chen}.

\section{Conclusions}
\label{sec:conclusion}
In this review 

\section{Conclusions}
\label{sec:conclusion}

In this article we have presented the MSSM3RHN lagrangian in terms of superfields. Then we presented the mass spectrum of neutrinos and sneutrinos with three 
right handed neutrinos and with only one right-handed neutrino and we can get neutrinos are Majorana particles or Dirac particles. At the end we 
presented some flat directions of these model.

We also present, very quickly, some others interestings supersymmetric models such as SUSYB-L, SUSYLR and two SUSYGUTS models.

We, also, showed how to generate masses to neutrinos in some interesting Supersymmetric models. All the models we reviwed here are intersting 
and can be tested at the new colliders machine. 

We hope this review can be useful to all the people wants to learn about Supersymmetry and to 
study the phenomenology in the neutrinos sector.

\begin{center}
{\bf Acknowledgments} 
\end{center}
The author would like to thanks to Instituto de F\'\i sica Te\'orica (IFT-Unesp) for their nice 
hospitality during the period I developed this review about SUSY.

\appendix

\section{Lagrangian}

The Lagrangian of this model is defined at Eq.(\ref{totallagmssm3rhn}). The ${\cal L}_{Quarks},{\cal L}_{Gauge}$ and ${\cal L}_{Higgs}$ are the same 
as in the MSSM and due it we do not write their expressions \cite{Baer:2006rs,dress,Rodriguez:2019mwf}.

\section{Lepton Lagrangian}
\label{a1}

The lagrangian defined at Eq.(\ref{supersymmetricpiece}), has the following components
\begin{eqnarray}
{\cal L}_{Lepton}=&=&{\cal L}^{lep}_{F}+ {\cal L}^{lep}_{D}+ {\cal L}^{lep}_{cin}+{\cal L}^{lep}_{l \tilde{l} \tilde{V}}.
\label{lint}
\end{eqnarray}
The first two terms in Eq.(\ref{lint}) are the usual $F$ terms given by:
\begin{eqnarray}
{\cal L}^{lep}_{F}&=& \sum_{i=1}^{3}\left( \vert F_{L_{iL}} \vert^{2}+ \vert F_{l_{iR}} \vert^{2}+ \vert F_{N_{iR}} \vert^{2} \right), 
\label{fterleptons}
\end{eqnarray}
while for the $D$ terms
\begin{eqnarray}
{\cal L}^{lep}_{D}&=& 
g \left[ \tilde{L}^{\dagger}_{iL} \left( \frac{\sigma^{a}}{2}\right)\tilde{L}_{iL} \right] D^{a}+
g^{\prime}\left[ \tilde{L}^{\dagger}_{iL}\left( \frac{-1}{2} \right)\tilde{L}_{iL} \right] D^{\prime}+
g^{\prime}\left[ \tilde{l}^{\dagger}_{iR}\left( \frac{2}{2} \right)\tilde{l}_{iR} \right] D^{\prime}, \nonumber \\ 
\label{dterleptons} 
\end{eqnarray}
where $\sigma^{a}$ (with $a=1,2,3$) are the generators of $SU(2)_{L}$.

The others terms in the lagrangian are given by
\begin{eqnarray}
{\cal L}^{lep}_{cin}&=&- \left[ \left( {\cal D}_{m}\tilde{L}_{i} \right)^{\dagger}\left( {\cal D}^{m}\tilde{L}_{iL} \right) +
\left( {\cal D}_{m}\tilde{l}_{iR} \right)^{\dagger}\left( {\cal D}^{m}\tilde{l}_{iR} \right) +
\left( \partial_{m}\tilde{N}_{iR} \right)^{\dagger}\left( \partial^{m}\tilde{N}_{iR} \right) \right. 
\nonumber \\ &+& \left.
\imath \bar{L}_{i}\bar{\sigma}^{m} \left( {\cal D}_{m}L_{i} \right) +
\imath \bar{l}_{iR}\bar{\sigma}^{m} \left( {\cal D}_{m}l_{iR} \right) +
\imath \bar{N}_{iR}\bar{\sigma}^{m} \left( \partial_{m}N_{iR} \right) \right], 
\label{cinlepton}
\end{eqnarray}
where the covariant derivatives are defined as
\begin{eqnarray}
{\cal D}_{m}\tilde{L}_{i}&=&\partial_{m}\tilde{L}_{i}+ 
\imath g\left[ \left(\frac{\sigma^{a}}{2}\right) V^{a}_{m}\right] \tilde{L}_{i}+ 
\imath g^{\prime} \left[ \left( \frac{-1}{2} \right) V_{m} \right] \tilde{L}_{i},  \nonumber \\
{\cal D}_{m}\tilde{l}_{iR}&=&\partial_{m}\tilde{l}_{iR}+ 
\imath g^{\prime}\left[ \left( \frac{2}{2}\right) V_{m} \right] \tilde{l}_{iR}.
\label{derivadascovariantesnosleptons}
\end{eqnarray}

The last term is
\begin{eqnarray}
{\cal L}^{lep}_{l \tilde{l} \tilde{V}}&=&- \imath g \sqrt{2}\left[
\bar{L}_{iL} \left(\frac{\sigma^{a}}{2}\lambda^{a}_{A}\right)\tilde{L}_{iL}-
\overline{\tilde{L}}_{iL} \left(\frac{\sigma^{a}}{2} \overline{\lambda^{a}_{A}}\right)L_{iL}  \right]
- \imath g^{\prime} \sqrt{2}\left[
\bar{L}_{iL} \left(\frac{-1}{2}\lambda_{B}\right)\tilde{L}_{iL}
\right. \nonumber \\ &-& \left.
\overline{\tilde{L}}_{iL} \left(\frac{-1}{2} \overline{\lambda_{B}}\right)L_{iL}  \right]
- \imath g^{\prime} \sqrt{2}\left[
\bar{l}_{iR} \left(\frac{2}{2}\lambda_{B}\right)\tilde{l}_{iR}-
\overline{\tilde{l}}_{iR} \left(\frac{2}{2} \overline{\lambda_{B}}\right)l_{iR}  \right].
\end{eqnarray}

Our superpotential $W$ is defined at Eq.(\ref{superpotentialmssm3rhnRPC}). We can write it in field components as
\begin{eqnarray}
W_{2}&=&{\cal L}^{W2}_{F}+{\cal L}^{W2}_{HMT}, \nonumber \\
W_{3}&=&{\cal L}^{W3}_{F}+{\cal L}^{W3}_{llH}+{\cal L}^{W3}_{l \tilde{l} \tilde{H}},
\end{eqnarray}
Where the $F$ terms are
\begin{eqnarray}
{\cal L}^{W2}_{F}&=& \mu_{H} \left( F_{H_{1}}H_{2}+F_{H_{2}}H_{1} \right) +hc,
\nonumber \\
{\cal L}^{W3}_{F}&=& f^{l}_{ij}\left( 
F_{H_{1}}\tilde{L}_{iL}\tilde{l}_{jR}+F_{L}H_{1}\tilde{l}_{jR}+F_{l}H_{1}\tilde{L}_{iL} \right) 
+f^{\nu}_{ij}\left( 
F_{H_{2}}\tilde{L}_{iL}\tilde{N}_{jR}+F_{L}H_{2}\tilde{N}_{jR}+F_{N}H_{2}\tilde{L}_{iL} \right) \nonumber \\
&+&hc, 
\label{fterms}
\end{eqnarray}
while the others parts
\begin{eqnarray}
{\cal L}^{W2}_{HMT}&=&- \mu_{H} \tilde{H}_{1}\tilde{H}_{2} + hc, \nonumber \\
{\cal L}^{W3}_{llH}&=&- \left[
f^{l}_{ij} \left( H_{1}L_{iL}\right)l_{jR}+ 
f^{\nu}_{i} \left( H_{2}L_{iL}\right) N_{iR}
\right] +hc. \nonumber \\
{\cal L}^{W3}_{l\tilde{l}\tilde{H}}&=&- \left\{
f^{l}_{ij} \left[ \left( \tilde{H_{1}}L_{iL} \right) \tilde{l}_{jR}+
\left( \tilde{H_{1}}\tilde{L}_{iL} \right) l_{jR} \right]
+ f^{\nu}_{ij} \left[ \left( \tilde{H_{2}}L_{iL} \right) \tilde{N}_{jR}+
\left( \tilde{H_{2}}\tilde{L}_{iL} \right) N_{jR} \right] +hc \right\} . \nonumber \\
\label{compsup}
\end{eqnarray}
Here we have omit the expansion to quarks.

\end{document}